\documentclass[fleqn,usenatbib]{mnras}

\usepackage{newtxtext,newtxmath}
\usepackage{mathrsfs}

\usepackage[T1]{fontenc}

\DeclareRobustCommand{\VAN}[3]{#2}
\let\VANthebibliography\thebibliography
\def\thebibliography{\DeclareRobustCommand{\VAN}[3]{##3}\VANthebibliography}

\usepackage{graphicx}	
\usepackage{amsmath}	

\newcommand{\nustar}{{\it NuSTAR}}
\newcommand{\swift}{{\it Swift}}
\newcommand{\xmm}{{\it XMM-Newton}}
\newcommand{\chandra}{{\it Chandra}}
\newcommand{\eps}{erg s$^{-1}$}
\newcommand{\ecs}{erg cm$^{-2}$ s$^{-1}$}
\newcommand{\pcm}{cm$^{-2}$}

\newcommand{\M}{$M_{\odot}$}

\newcommand{\nhl}{$N_{\rm H}^{\rm los}$}
\newcommand{\nht}{$N_{\rm H}^{\rm tor}$}

\newcommand{\borus}{{\tt borus02}}

\title[Low-Accreting AGNs]{Coronal Properties of Low-Accreting AGNs using Swift, XMM-Newton and NuSTAR Observations}

\author[A. Jana et al.]{
Arghajit Jana$^1$\thanks{argha0004@gmail.com}, 
Arka Chatterjee$^2$\thanks{arka019icsp@gmail.com},
Hsiang-Kuang Chang$^1$,
Prantik Nandi$^3$,
Rubinur K.$^{4,5}$,
\newauthor
Neeraj Kumari$^{3}$,
Sachindra Naik$^3$,
Samar Safi-Harb$^2$,
Claudio Ricci$^{6,7}$
\\
$^{1}$ Institute of Astronomy, National Tsing Hua University, Hsinchu 300044, Taiwan\\
$^2$ Department of Physics and Astronomy, University of Manitoba, Winnipeg, MB R3T 2N2, Canada\\
$^3$ Astronomy \& Astrophysics Division, Physical Research Laboratory, Navrangpura, Ahmedabad, Gujarat 38009, India \\
$^4$ National Centre for Radio Astrophysics - Tata Institute of Fundamental Research (NCRA-TIFR), S. P. Pune University Campus, Ganeshkhind, Pune 411007, India \\
$^5$ Institute of Theoretical Astrophysics, University of Oslo, P.O box 1029 Blindern, 0315 OSLO, Norway\\
$^6$ N{\'u}cleo de Astronom{\'i}a de la Facultad de Ingenier{\'i}a, Universidad Diego Portales, Av. Ej{\'e}rcito Libertador 441, Santiago, Chile\\
$^{7}$ Kavli Institute for Astronomy and Astrophysics, Peking University, Beijing 100871, People's Republic of China\\
}

\date{Accepted XXX. Received YYY; in original form ZZZ}

\pubyear{2023}

\begin{document}
\label{firstpage}
\pagerange{\pageref{firstpage}--\pageref{lastpage}}
\maketitle

\begin{abstract}
We studied the broadband X-ray spectra of {\it Swift}/BAT selected low-accreting AGNs using the observations from {\it XMM-Newton}, {\it Swift}, and {\it NuSTAR} in the energy range of $0.5-150$~keV. Our sample consists of 30 AGNs with Eddington ratio, $\lambda_{\rm Edd}<10^{-3}$. We extracted several coronal parameters from the spectral modelling, such as the photon index, hot electron plasma temperature, cutoff energy, and optical depth. We tested whether there exists any correlation/anti-correlation among different spectral parameters. We observe that the relation of hot electron temperature with the cutoff energy in the low accretion domain is similar to what is observed in the high accretion domain. We did not observe any correlation between the Eddington ratio and the photon index. We studied the compactness-temperature diagram and found that the cooling process for extremely low-accreting AGNs is complex. The jet luminosity is calculated from the radio flux, and observed to be related to the bolometric luminosity as $L_{\rm jet} \propto L_{\rm bol}^{0.7}$, which is consistent with the standard radio-X-ray correlation.
\end{abstract}

\begin{keywords}
galaxies: active -- galaxies: Seyfert -- X-rays: galaxies -- galaxies: quasars: supermassive black holes -- accretion: accretion discs -- black hole physics
\end{keywords}

\section{Introduction}
\label{sec:intro}
Active galactic nuclei (AGNs) are powered by the accreting supermassive black holes (SMBHs) that reside at the centre of most galaxies \citep{Rees1984}. The matter get accreted onto the SMBH, where the gravitational potential energy is converted into radiation, which is emitted over the entire electromagnetic spectrum. The X-rays are thought to be produced in a hot electron cloud, known as the corona, located in the vicinity of the black hole \citep{HM1991,Naayan1994,CT1995,Done2007}. The primary X-ray continuum is produced through the inverse-Comptonization \citep{ST80,ST1985,HM1991} of the seed UV photons from the standard accretion disc \citep{SS1973}. The X-ray continuum can be reprocessed by the accretion disc and/or the molecular torus, which produces a reflection hump at $\sim 15-40$~keV and an iron K$\alpha$ line at $\sim 6.4$~keV \citep{George1991,Matt1991}. Additionally, an excess in the soft X-ray energy band ($<2$~keV), known as soft-excess, is observed in several sources \citep{Singh1985,Arnaud1985}. The origin of the soft-excess is still debated, and some of the possible explanations proposed include blurred reflection from the inner disc \citep{lohfink2012}, a warm corona \citep{Mehdipour2011,Done2012}, or a small number of scattering in a hot corona \citep{Nandi2021}.

In general, an AGN is classified as a low-luminosity AGN (LLAGN), if the bolometric luminosity is $L_{\rm bol}<10^{44}$ \eps \citep[e.g.,][]{Gu2009}. 
Recent studies have suggested that the mass-normalized accretion rate is the primary driver in the evolution of the circumnuclear gas in AGNs \citep[e.g.,][]{Ricci2017bat}. It is believed that the accretion mechanism is different in the low-accreting AGNs (LAC-AGNs; Eddington ratio, $\lambda_{\rm Edd} = L_{\rm bol}/L_{\rm Edd} <10^{-3}$, where $L_{\rm Edd}$ is Eddington luminosity) from the high-accreting AGNs \citep[HAC-AGNs; $\lambda_{\rm Edd} >10^{-3}$; e.g.,][]{Ho2009,Yang2015,Kawamuro2016}.

Theory predicts that if the accretion rate falls below a critical level, the inner accretion flow changes from the geometrically thin, optically thick accretion disc \citep{SS1973} to an optically thin, radiatively inefficient accretion flow \citep{Esin1997}. The correlation between the $\lambda_{\rm Edd}$ and photon index ($\Gamma$) could be considered as an observational manifestation of such a theoretical claim. While a positive correlation has been found for high-luminosity AGNs \citep[e.g.,][]{Shemmer2006,Shemmer2008}, an anti-correlation is observed for low-luminosity AGNs \citep[e.g.,][]{Ho2009,Gu2009}. It is, therefore, widely believed that in LLAGNs, the standard thin accretion disc is replaced by a radiatively-inefficient accretion flow \citep[e.g.,][]{Narayan1998,Quataert2001,Ho2009}. The absence of the big-blue bump in the SED of these objects suggests that the thin accretion disc gets truncated at a large distance from the BH \citep[e.g.,][]{Mason2012,Nemen2014}. Moreover, most LLAGNs do not show the broad iron K-line feature, suggesting that the standard disc does not get extended to the innermost region around the SMBH \citep[e.g.,][]{Kawamuro2016,Younes2019}.

In the case of the HAC-AGNs, the majority of the seed photons that produce the X-ray emission, are most likely thermal and originate in a standard accretion disc \citep[e.g.,][]{SS1973,Malkan1982}. However, for LAC-AGNs, the origin of the seed photons could be dominated by non-thermal processes, such as synchrotron emission occurring in the jet or within the corona \citep[e.g.,][]{Yang2015}. The size, shape, and geometry of the corona are highly debated. Various studies suggest that the X-ray corona is compact \citep[$\sim 10~R_g$, where $R_g$ is gravitational radius; e.g.,][]{McHardy2005,Chartas2009,Risaliti2011,Reis2013,Uttley2014} and located close to the black hole \citep[$\sim 3-10~R_g$; e.g.,][]{Fabian2009,Kara2013,Zoghbi2012,Fabian2015,Fabian2017}. In the low-accreting AGNs \citep[$L < 10^{-4} L_{\rm Edd}$, where $L_{\rm Edd}$ is the Eddington luminosity; e.g.,][]{Reis2013}, the hard X-rays could originate in a hot quasi-spherical accretion flow or in an extended corona ($\sim 100~R_g$).

The corona in AGN is typically characterized by the electron plasma temperature ($kT_{\rm e}$) and optical depth ($\tau_{\rm e}$). The electron temperature is directly related to the cutoff energy ($E_{\rm cut}$), while the photon index is connected to both $kT_{\rm e}$ and $\tau_{\rm e}$ \citep[e.g.,][]{ST80,PSS1983,Petrucci2001}. The photon index ($\Gamma$) and the high energy cutoff ($E_{\rm cut}$) can be inferred by X-ray spectroscopic analysis of AGN. The photon index has been studied over the past three decades. In contrast, the study of the high energy cutoff has been limited until more recent times due to the restricted bandpass of the X-ray facilities. Recently, the cutoff energy of the AGNs has been measured using various observatories with hard X-ray instruments, e.g., \textit{BeppoSAX} \citep{Dadina2007,Perola2002}, \textit{INTEGRAL} \citep{deRosa2012,Molina2009,Molina2013,Panessa2011}, \nustar~ \citep{Balokovic2020,Kamraj2018,Kamraj2022,Rani2019} and \swift~ \citep{Ricci2017bat,Ricci2018,Trakhtenbrot2017}. In general, the cutoff energy is observed in a wide range of $\sim 50-500$~keV \citep[e.g.,][]{Ricci2017bat,Ricci2018,Balokovic2020}. \citet{Ricci2018} analyzed 838 BAT AGNs and found that the median of the cutoff energy, $E_{\rm cut} = 160 \pm 41$~keV for $L/L_{\rm Edd}>0.1$, and $E_{\rm cut} = 370 \pm 51$~keV for $L/L_{\rm Edd}<0.1$. They also found the median of the hot electron temperature and optical depth as $kT_{\rm e} = 105 \pm 18$~keV and $\tau_{\rm e} = 0.25 \pm 0.06$, respectively. 

While the coronal properties of the HAC-AGNs have been explored extensively in past \citep[e.g.,][]{Ricci2017bat,Ricci2018,Hinkle2021,Kamraj2018,Kamraj2022,Balokovic2020,Rani2019}, the low accretion domain has been considerably less explored \citep[e.g.,][]{Younes2019}. Several X-ray studies of LLAGNs have been performed to study the variation of the photon index \citep[e.g.,][]{Kawamuro2016,Ho2009,Gu2009,Shemmer2008}.

In the present paper, we study the coronal properties of low-accreting AGNs using broadband X-ray data obtained from \nustar~ and \swift/BAT in the $3-150$~keV range. Using Comptonization models, we constrain the main coronal parameters and study possible trends among them. The paper is organized in the following way. First, in \S2, we describe the sample selection and data reduction processes. Then, we discuss the analysis procedure in \S3. Next, the results of our work are presented and discussed in \S4. Finally, we summarize our findings in \S5.

\begin{table*}
\caption{Information on the selected sources}
\label{tab:info}
\centering
\begin{tabular}{lccccccccccc}
\hline
 & Name & Swift Name & Type & R.A. (J2000) & Decl.(J2000) & Redshift & $\log(M_{\rm BH}$) & Ref. \\
\hline
(1) & NGC 454E  &J0114.4--5522& Seyfert 2     & 18.575  & --55.401  &0.0121 &$8.52\pm0.45$ & 1 \\
(2) & NGC 1052 & J0241.3--0816 & Seyfert 2    & 40.270 & --8.256    &0.005  & $8.96\pm 0.29$ & 1 \\
(3) & NGC 2110 & J0552.2--0727 & Seyfert 2    & 88.046 & --7.457    & 0.007 & 9.38           & 2 \\ 
(4) & NGC 2655	&J0856.0+7812 &	Seyfert 2	  & 133.90  &  78.20	&0.0047	&$7.70\pm0.20$ & 3\\
(5) & NGC 3079  &J1001.7+5543 & Seyfert 2     & 150.49  &  55.679   &0.0037 &$8.27\pm0.30$& 1 \\
(6) & NGC 3147	&J1017.8+7340 &	Seyfert 2	  & 155.45  &  73.41	&0.0093	&8.79 & 4\\
(7) & NGC 3718	&J1132.7+5301 &	LINER 1.9     & 173.22  &  53.02	&0.0033	&9.53 & 1 \\
(8) & NGC 3786	&J1139.5--6526&Seyfert 1.9    & 174.94  &  31.96	&0.0089 &7.53 & 5 \\
(9) & NGC 3998	&J1157.8+5529 &	LINER 1.9     & 179.48  &  55.45    &0.0035	&$9.93\pm0.33$& 1 \\
(10) & NGC 4102	&J1206.2+5243 &	LINER	      & 181.59  &  52.71	&0.0028	&$8.75\pm0.33$ & 1 \\
(11)& NGC 4258	&J1219.4+4720 &	Seyfert 1.9/LINER & 184.75  & 47.29 &0.0015&$7.57\pm0.35$ & 1\\
(12)& NGC 4579	&J1237.5+1182 &	LINER 1.9     & 189.38  &  11.82	&0.0051	&8.10 & 6\\
(13)& NGC 5033  &J1313.6+3650B &Seyfert 1.5   & 198.406 &  36.826   &0.0029 &$7.86\pm0.35$ & 1\\
(14)& NGC 5283 & J1341.5+6742   &  Seyfert 2   & 205.299 & 67.691    & 0.010 & $8.87+0.30$   & 1 \\
(15)& NGC 5290 & J1345.5+4139 & Seyfert 2    & 206.329 & 41.713    &0.0080 & $7.76\pm 0.31$ & 1 \\
(16)& NGC 5899  &J1515.0+4205 &Seyfert 2      & 228.788 &  42.063   &0.0080 &$8.66\pm0.31$ &1 \\
(17)& NGC 6232  &J1643.2+7036 &Seyfert 1      & 250.721 &  70.643   &0.0148 &$7.43\pm0.52$& 1 \\
(18)& NGC 7213	&J2209.4--4711& Seyfert 1     & 332.33  & --47.17	&0.0058	&7.99 & 5 \\
(19)& NGC 7674  & J2328.1+0883 & Seyfert 2    & 352.031 & 8.835     & 0.028 & 9.18 & 2 \\
(20)& Mrk 18    &J0902.0+6007 &Seyfert 2      & 135.493 & 60.152    &0.0111 & $7.85\pm 0.30$ & 1 \\
(21)& Mrk 273   &J1344.7+5588 & Seyfert 2      & 206.175& 55.887     &0.0379 & $9.02\pm 0.04$ & 7 \\
(22)& ARP 102B	&J1719.7+4900 &	Seyfert 1      & 259.81  &  48.98	&0.0242	&$8.92 \pm0.34$  & 1\\
(23)& ESO 253--003&J0525.3-4600 & Seyfert 2   & 81.381 	&--45.965   & 0.042 & 9.84  & 2 \\
(24)& ESO 506--027&J1238.9--2720 &  Seyfert 2  & 189.722  &--27.294   & 0.025 & $8.99\pm0.29$ &1\\
(25)& HE 1136--2304 &J1139.0--2323& Seyfert 1.9&74.713 & --23.360    &0.027  & 9.39 & 2 \\
(26)& IGR J11366--6002 & J1136.7+6738 & Seyfert 1& 174.104 & 67.645    & 0.014 & 8.56 & 2 \\
(27) & IC 4518A &J1457.8--4308 &Seyfert 2 & 224.460 & --43.116 & 0.016 & 8.79 & 2 \\
(28)& UGC 12282 & J2258.9+4054  &  Seyfert 1   & 344.696 & 40.918    &0.017  & $9.80\pm0.35$ & 1\\
(29)& LEDA 214543& J1650.5+0434 &  Seyfert 2   & 252.656 & 4.620     &0.032  & $9.83\pm0.32$ & 1 \\
(30)& Z367--9   & J1621.2+8104  &  Seyfert 2   & 244.927 & 81.062    & 0.027 & $9.82\pm0.32$ & 1 \\
\hline
\end{tabular}
\leftline{(1) \citet{Koss2017}, (2) \citet{}{Koss2022}, (3) \citet{Tully1988}, (4) \citet{Merloni2003} , (5) \citet{Woo2002}, (6) \citet{Younes2019}, (7) \citet{U2013}.}
\end{table*}

\section{Sample and Data Reduction}
\label{sec:sample}

\subsection{Sample Selection}
\label{sec:sam-sel}
The primary aim of this work is to investigate the coronal properties of low-accreting AGNs. We chose our sample from the all-sky \swift/BAT hard X-ray survey\footnote{\url{https://swift.gsfc.nasa.gov/results/bs105mon/}} \citep{Oh2018}. The BAT survey was compiled for 105 months, and the spectra are stacked together. The survey has a sensitivity of $8.4 \times 10^{-12}$ \ecs~ in $14-195$~keV range, and is almost unbiased by obscuration up to $N_{\rm H}\sim 10^{24}\rm\,cm^{-2}$ \citep{Ricci:2015}. Initially, we chose a sample of AGNs within the range of $14-195$~keV, where the luminosity is less than $10^{45}$ \eps. Our main goal is to select sources with an Eddington ratio $<10^{-3}$.

\begin{figure*}
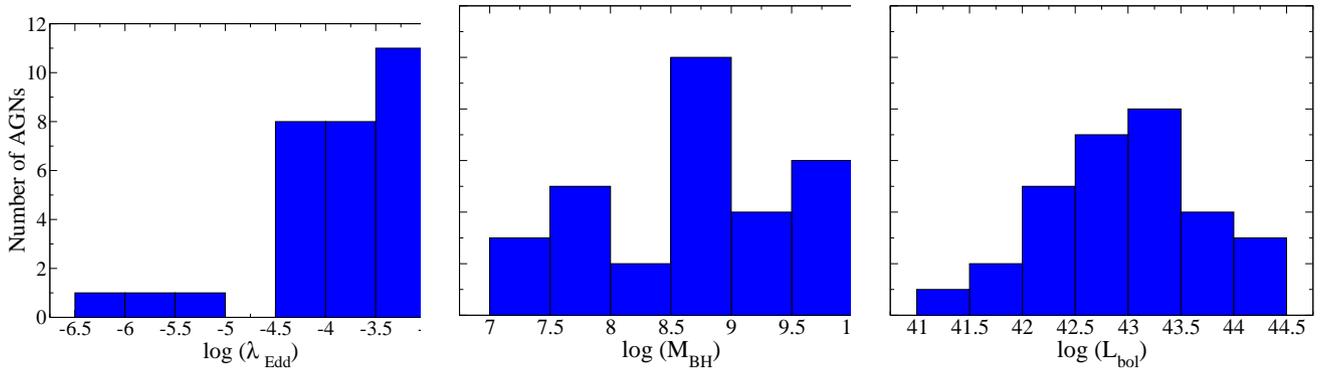

\centering
\includegraphics[width=5.85cm]{edd-hist.eps}
\includegraphics[width=5.6cm]{mbh-hist.eps}
\includegraphics[width=5.6cm]{bol-hist.eps}
\caption{Histograms of Eddington ratio ($\lambda_{\rm Edd}$), super-massive black hole mass ($M_{\rm BH}$), and bolometric luminosity ($L_{\rm bol}$) are shown in the left, middle, and right panels, respectively.}
\label{fig:hist1}
\end{figure*}

We selected all BAT AGNs with publicly available \nustar~ observations. We also added \xmm~ or \swift/XRT observations for the soft X-ray band ($0.5-10$ keV) to construct the spectra in a broad energy range of $0.5-150$~keV\footnote{Check Appendix~A for details.}. From the combined spectra, we performed spectral analysis to calculate the bolometric luminosity ($L_{\rm bol}$) and Eddington ratio ($\lambda_{\rm Edd}$; see Section~\ref{sec:analysis} for details). We selected sources with $\lambda_{\rm Edd} < 10^{-3}$ for our sample. We excluded several sources from our sample (with $\lambda_{\rm Edd}<10^{-3}$) due to low signal-to-noise ratio (SNR), e.g., NCG~660, NGC~3486, NGC~678, or the presence of other sources in the field of view, e.g., NGC~5194. In the case of NGC~5194, several ULXs have been detected in the \nustar~ field. As \nustar~ does not resolve them, the emission is not purely from the AGN \citep{Brightman2018}. Hence, we did not include NGC~5194 in our sample. The final sample consists of 30 sources and are tabulated in Table~\ref{tab:info}. In Figure~\ref{fig:hist1}, we show the distribution of Eddington ratio ($\lambda_{\rm Edd}$), black hole mass ($M_{\rm BH}$), and bolometric luminosity ($L_{\rm bol}$) of our sample, in the left, middle, and right panels, respectively. Our sample extends over three orders of magnitude for all three parameters.

\subsection{Data Reduction}
\label{sec:data}

\subsubsection{NuSTAR}
\label{sec:nustar}
\nustar~ is a hard X-ray focusing telescope with two identical modules, FPMA and FPMB, and operates in the $3-78$~keV energy range \citep{Harrison2013}. We obtained \nustar~ data from NASA's HEASARC archive\footnote{\url{https://heasarc.gsfc.nasa.gov/cgi-bin/W3Browse/w3browse.pl}}. The data were reprocessed with the \nustar~ Data Analysis Software ({\tt NuSTARDAS}\footnote{\url{https://heasarc.gsfc.nasa.gov/docs/nustar/analysis/}}, version 1.4.1). We generated clean event files with the {\tt nupipeline} task, using standard filtering criteria. The data were calibrated using the latest calibration data files available in the NuSTAR calibration database\footnote{\url{http://heasarc.gsfc.nasa.gov/FTP/caldb/data/nustar/fpm/}}. The source and background products were extracted by considering circular regions with 60\,arcsec, and 90\,arcsec radii, centered at the source coordinates and away from the source, respectively. The spectra were extracted using the {\tt nuproduct} task and then rebinned to ensure that they had at least 20 counts per bin by using the {\tt grppha} task. For each source, we used the \nustar~ observation with the longest exposure, except for NGC~3718 for which we co-added the spectra from four continuous observations to improve the SNR using the {\tt FTOOL} task \textsc{addascaspec}.

\subsection{Swift}
\label{sec:swift}
The $0.5-8$ keV \swift/XRT spectra were generated using the standard online tools provided by the UK Swift Science Data Centre \citep{Evans2009}\footnote{\url{https://www.swift.ac.uk/user_objects/}}. We utilized the \swift/XRT spectra for 17 objects when the simultaneous observations were available with \nustar. For five objects, we stacked several XRT spectra together to achieve a good SNR.

The $14-150$~keV \swift/BAT spectra and response matrices were obtained from the 105-month Swift-BAT All-sky Hard X-Ray Survey\footnote{\url{https://swift.gsfc.nasa.gov/results/bs105mon/}}. 

\subsubsection{XMM-Newton}
\label{sec:xmm}
We used {\it XMM-Newton} \citep{Jansen2001} EPIC/PN observations in the $0.5-10$ keV energy range in our analysis. The data files were reduced using the Standard Analysis Software (SAS) version 20.0.0. The raw PN event files were processed using \textsc{epchain} task. We checked for particle background flare in the $10-12$ keV energy range. The Good Time Interval file was generated using the \texttt{SAS} task \textsc{tabgtigen}. The source and background spectra were extracted from a circular region of 30$\arcsec$ centred at the position of the optical counterpart and from a circular region of 30$\arcsec$ radius away from the source, respectively. The background region is selected in the same CCD where no other X-ray sources are present. Using \textsc{especget} task, we generated the source and background spectra. We checked for pileup using the \textsc{epatplot} task. We did not find any source that suffered from the pileup.

We used \xmm~ spectra for 13 objects. For eight sources, the \xmm~ observations were made simultaneously with the \nustar. For the rest five sources, we used non-simultaneous observations. For the non-simultaneous observations with \xmm, and \nustar, we checked for spectral variability. The spectral variability is presented in detail in Section~\ref{sec:spec-var}. The detailed observation log is tabulated in Table~\ref{tab:log}.



\begin{table*}
\caption{Observation Log}
\label{tab:log}
\centering
\begin{tabular}{lcccccccc}
\hline
Object & NuSTAR ID & Date & Exp & XMM-Newton  & Date & Exp \\
 &  & (yyyy-dd-mm) & (ks) & or XRT ID &(yyyy-dd-mm) & (ks)\\
\hline
NGC 454E  & 60061009002&    2016-02-14& 24  & 00080016001 & 2016-02-14 & 6  \\
NGC 1052  & 60061027002&    2013-02-14& 16  & 0790980101$^{\rm X*}$  & 2017-01-17 & 71 \\
NGC 2110  & 60061061002&    2012-10-05& 15  & 0145670101$^{\rm X*}$  & 2003-03-05 & 60 \\
NGC 2655  & 60160341004&	2016-11-10&	16  & 00081037001--02 & 2016-11-02 -- 03 & 7 \\
NGC 3079  & 60061097002&    2013-11-12& 22  & 00080030001 & 2013-11-12 & 7 \\
NGC 3147  & 60101032002&	2015-12-27&	49  & 0405020601$^{\rm X*}$  & 2006-06-10 & 18 \\
NGC 3718  & 60301031002&	2017-10-24&	25  & 0795730101$^{\rm X}$  & 2017-10-24 & 38\\
  & 60301031004&	2017-10-27&	90  \\
  & 60301031006&	2017-10-30&	57  \\
  & 60301031008&	2017-11-03&	57  \\
NGC 3786  & 60061349002&	2014-06-09&	22  & 00080684001 & 2014-06-09 & 4\\
NGC 3998  & 60201050002&	2016-10-25&	104 & 0790840101$^{\rm X}$  & 2016-10-26 & 25 \\
NGC 4102  & 60160472002&	2015-11-19&	21  & 00081110001 & 2015-11-09 & 7 \\
NGC 4258  & 60101046002&	2015-11-16&	55  & 00081700001 & 2015-11-16 & 2\\
NGC 4579  & 60201051002&	2016-12-06&	117 & 0790840201$^{\rm X}$  & 2016-12-06 & 23 \\
NGC 5033  & 60601023002&    2020-12-08& 104 & 0871020101$^{\rm X}$  & 2020-12-10 & 21 \\
NGC 5283  & 60465006002&    2018-11-17& 33  & 00088264001 & 2018-11-17 & 7 \\
NGC 5290  & 60160554002&    2021-07-28& 19  &  00011388002-- & 2019-05-07 & 9\\
          &            &               &     & 00011388007 & to 2020-05-26 \\
NGC 5899  & 60061348002&    2014-04-08& 24  & 00080683001 & 2014-04-08 & 7 \\
NGC 6232  & 60061328002&    2013-08-17& 18  & 00080537001--02 & 2013-08-17 -- 18 & 7\\
NGC 7213  & 60001031002&	2014-10-05&	102 & 00080811001 & 2014-10-06 & 2\\
NGC 7674  & 60001151002&    2014-09-30& 52  & 0200660101$^{\rm X*}$  & 2004-06-02 & 10 \\
Mrk 18    & 60061088002&    2013-12-15& 20  & 00080406001 & 2013-12-15 & 7 \\
Mrk 273   & 60002028002&    2013-11-04& 70  & 0722610201$^{\rm X}$  & 2013-11-04 & 23\\
ARP 102B  & 60160662002&	2015-11-24&	22  & 00081204001 & 2015-11-24 & 7\\
ESO 253--003 & 60101014002 &2015-08-21& 23  & 0762920501$^{\rm X}$  & 2015-08-19 & 27 \\
ESO 506--027&60469006002&    2019-06-26& 19 & 0312191801$^{\rm X*}$  & 2006-01-24 & 12\\
HE 1136--2304 & 80002031003 & 2014-07-02 & 64& 0741260101$^{\rm X}$ & 2014-07-02 & 110 \\
IC 4518A & 60061260002      & 2013-08-02 & 8 & 00080141001& 2013-08-02 & 7 \\
IGC J11366--6002 & 60061213002 & 2014-10-29 & 22 & 00080058001--02 & 2014-10-29 -- 30 & 7 \\
UGC 12282  &60160812002&    2019-11-18& 29  & 00081292001 & 2019-11-18 & 7 \\
LEDA 96373 &60061073002& 2014-07-31 & 22 & 00080382001 & 2014-07-31 & 4 \\
LEDA 214543&60061273002&    2017-02-06& 21  & 00080172001 & 2017-02-06 & 6 \\
Z 367--9   &60061270002&    2014-12-21& 30  & 00080158001-- & 2014-09-22 & 13\\
           &           &               &    & 00080158002 & 2014-12-21 & \\
\hline
\end{tabular}
\leftline{$^{\rm X}$ mark the \xmm~ observations. $^{*}$ indicate non-simultaneous observations of \nustar~ and \xmm.}
\end{table*}

\begin{figure}
\centering
\includegraphics[width=8.5cm]{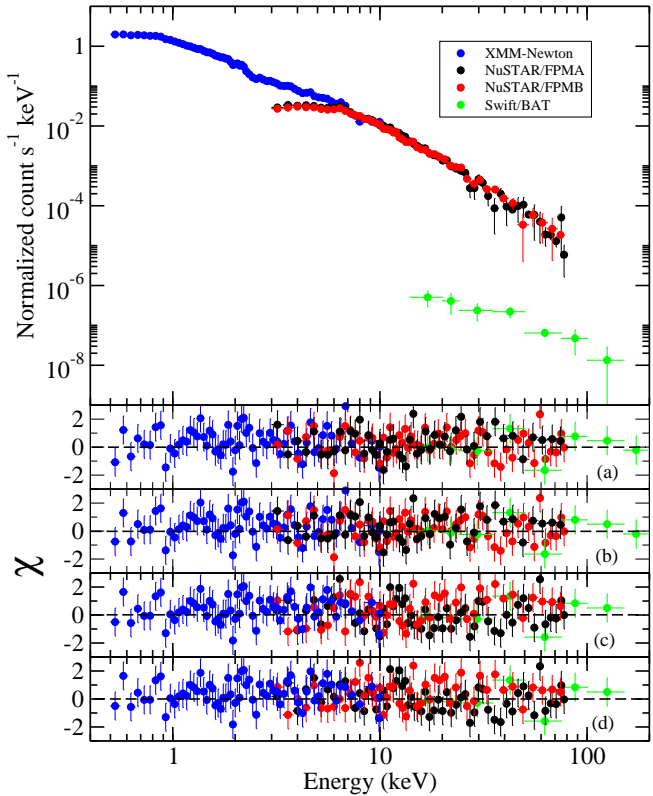}
\caption{Top panel: Representative spectrum of NGC~4579. The black, red, and green circles represent the data from the FPMA, FPMB, and BAT instruments, respectively. The distributions of $\chi$ are shown in the middle and bottom panels, obtained from fitting the data with (a) Model-1a, (b) Model-1b, (c) Model-2a, and Model-2d, respectively. }
\label{fig:spec}
\end{figure}

\begin{figure*}
\centering
\includegraphics[angle=270,width=5.9cm]{454.eps}
\includegraphics[angle=270,width=5.9cm]{3147.eps}
\includegraphics[angle=270,width=5.9cm]{3998.eps}\hspace{1cm}
\includegraphics[angle=270,width=5.9cm]{4102.eps}
\includegraphics[angle=270,width=5.9cm]{7213.eps}
\includegraphics[angle=270,width=5.9cm]{1136.eps}
\caption{Confidence contours between the photon index ($\Gamma$) and cutoff energy ($E_{\rm cut}$) are shown for NGC~454E (top left), NGC~3147 (top middle), NGC~3998 (top right), NGC~4102 (bottom left), NGC~7213 (bottom middle), and HE~1136--2304 (bottom right). The red, green, and blue contours represent 1~$\sigma$, $2~\sigma$, and 3~$\sigma$ levels, respectively.}
\label{fig:contour1}
\end{figure*}

\begin{figure*}
\centering
\includegraphics[angle=270,width=5.9cm]{nth-454.eps}
\includegraphics[angle=270,width=5.9cm]{nth-3147.eps}
\includegraphics[angle=270,width=5.9cm]{nth-3998.eps}
\hspace{1cm}
\includegraphics[angle=270,width=5.9cm]{nth-4102.eps}
\includegraphics[angle=270,width=5.9cm]{nth-7213.eps}
\includegraphics[angle=270,width=5.9cm]{nth-1136.eps}
\caption{Confidence contours between the photon index ($\Gamma$) and hot electron temperature ($kT_{\rm e}$) are shown for NGC~454E (top left), NGC~3147 (top middle), NGC~3998 (top right), NGC~4102 (bottom left), NGC~7213 (bottom middle), and HE~1136--2304 (bottom right). The red, green, and blue contours represent 1~$\sigma$, $2~\sigma$, and 3~$\sigma$ levels, respectively.}
\label{fig:contour2}
\end{figure*}

\begin{figure*}
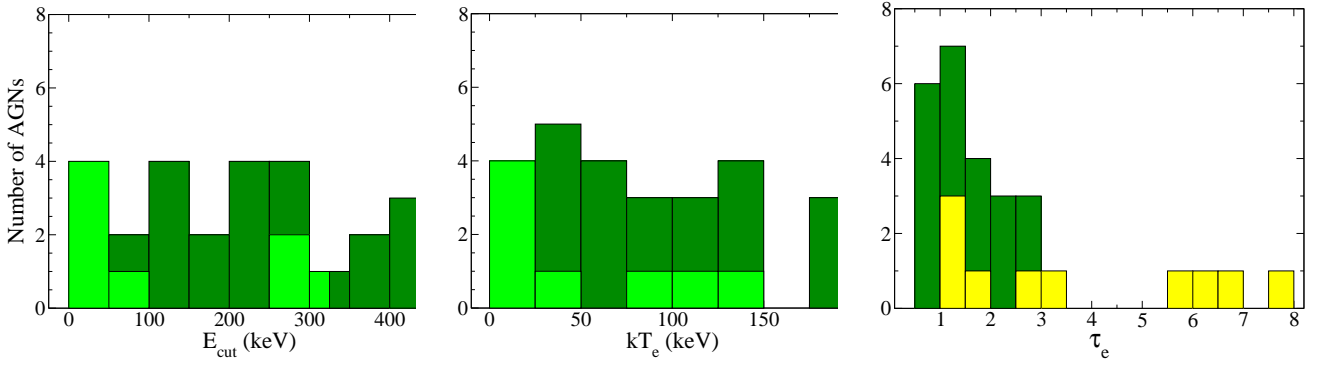

\centering
\includegraphics[width=5.85cm]{cut-hist.eps}
\includegraphics[width=5.5cm]{kt-hist.eps}
\includegraphics[width=5.6cm]{tau-hist.eps}
\caption{Histograms of cutoff energy ($E_{\rm cut}$), hot electron plasma temperature ($kT_{\rm e}$), and optical depth ($\tau_{\rm e}$) are shown in the left, middle, and right panels, respectively. The dark green bars represent the constrained parameters. The light green and yellow bars represent the lower limit and upper limit of the parameters, respectively.}
\label{fig:hist2}
\end{figure*}

\begin{figure}
\centering
\includegraphics[width=8.5cm]{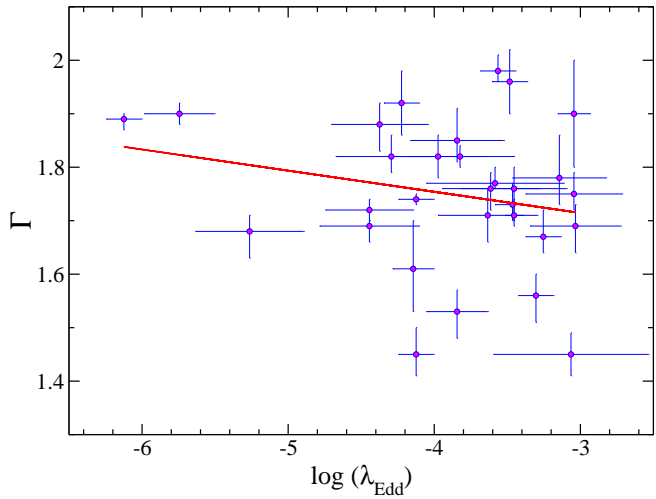}
\caption{The photon index ($\Gamma$) is plotted as a function of the Eddington ratio ($\lambda_{\rm Edd}$). The linear regression analysis gives $\Gamma = (-0.04\pm 0.03)\log \lambda_{\rm Edd} + (1.59\pm 0.13)$. The red line represents the best-fit of the linear regression analysis.}
\label{fig:cor-edd}
\end{figure}

\begin{figure}
\centering
\includegraphics[width=8.5cm]{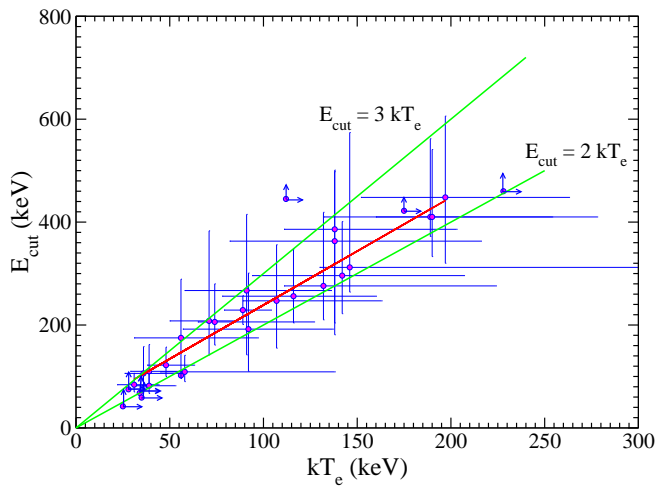}
\caption{Relation between the hot electron temperature ($kT_{\rm e}$) and cutoff energy ($E_{\rm cut}$) is shown. The solid red line represents the best fit to the data. The best fit is $E_{\rm cut}=(2.10\pm0.12) kT_{\rm e} + (29.4\pm12.1)$. Two green solid lines represent the relation $E_{\rm cut} = 2 kT_{\rm e}$ and $E_{\rm cut} = 3 kT_{\rm e}$. }
\label{fig:cor-kt}
\end{figure}

\section{Analysis}
\label{sec:analysis}

\subsection{Spectral Analysis}
\label{sec:spec}

The spectral analysis of combined spectra in the $0.5-150$~keV range was performed in \textsc{xspec} v12.10 \citep{Arnaud1996}. For our analysis, we adopted the cross-section from \citet{Verner1996}, and \textsc{angr} abundances \citep{Anders1989}. We used cross-normalization constants between the FPMA, FPMB, and BAT \citep{Madsen2015,Madsen2017} instruments while carrying out simultaneous spectral fitting. 

We started X-ray spectral modelling using an absorbed power law model with a cutoff at high energy. In \textsc{xspec}, the model reads as \textsc{zphabs*zcutoff}. We also added another component in the model for the absorption due to the Compton scattering, modelled with \textsc{cabs}. A component for the scattered primary emission, modelled with \textsc{constant*zcutoff} was added \citep[e.g.,][]{Gupta:2021}. For the reprocessed emission, we used the convolution model \textsc{reflect}\footnote{\url{https://heasarc.gsfc.nasa.gov/xanadu/xspec/manual/node297.html}} \citep{Magdziarz1995}. \textsc{Reflect} is a generalization of the widely used \textsc{pexrav} model. It describes the reflection from a cold, neutral semi-infinite slab. The model parameters are reflection fraction ($R$), inclination angle ($i$), iron abundance ($A_{\rm Fe}$), and metal abundances ($A_{\rm M}$). We also added a Gaussian function at $\sim 6.4$~keV to incorporate the iron K-line emission. For the soft-excess emission, we added a \textsc{blackbody} model. However, one can also use the \textsc{powerlaw} to approximate the soft-excess \citep[e.g.,][]{Nandi2021}. The model setup (hereafter Model-1a) reads in {\tt XSPEC} as,

\textsc{Const1*phabs1*(zphabs2*cabs*reflect*zcutoffpl1 + Gauss + const2*cutoffpl2 + blackbody)}.

where, \textsc{phabs1} represents the Galactic absorption and is calculated using \textsc{NH}\footnote{\url{https://heasarc.gsfc.nasa.gov/cgi-bin/Tools/w3nh/w3nh.pl}} tools of {\tt FTOOLs} \citep{HI4PI2016}. \textsc{Const1} represents the cross-normalization factor between the FPMA, FPMB, and BAT. The \textsc{zphabs2*cabs*zcutoffpl1} represents the absorbed direct primary emission. \textsc{const2*cutoffpl} represents the scattered primary emission, while \textsc{const2} is the scattering fraction ($f_{\rm Scat}$). The photon index ($\Gamma$), cutoff energy ($E_{\rm cut}$), normalization of \textsc{cutoffpl1} and \textsc{zcutoffpl2} are linked together. The column densities of the \textsc{cabs} and \textsc{zphabs2} models are tied together and represent the line-of-sight obscuration towards the AGN. We fixed the $A_{\rm Fe}$ and $A_{\rm M}$ at the Solar values, i.e., 1 and the inclination angle at 60\textdegree~ in our analysis. We allowed the Gaussian parameters to vary freely. However, when we could not constrain, we fixed the line energy at 6.4 keV and line width at 0.01, 0.05, or 0.1 keV, depending on the initial fitting.

We obtained good fits for all the sources using Model-1a. We noticed that the soft-excess is present in seven sources, namely, NGC~2655, NGC~4102, NGC~4258, NGC~5033, NGC~5290, NGC~7213, and HE~1136--2304. The spectra of the rest 23 sources can be fitted without the \textsc{blackbody} component in Model-1a. The scattered emission is present in 15 sources in our sample. The photon index ($\Gamma$) and cutoff energy ($E_{\rm cut}$) were obtained from the fitting. The hot electron temperature of the corona ($kT_{\rm e}$) can be calculated from the cutoff energy, using the empirical relation $E_{\rm cut} = 2 kT_{\rm e}$ (for $\tau_{\rm e} < 1$) or $E_{\rm cut} = 3 kT_{\rm e}$ (for $\tau_{\rm e} >> 1$) \citep{Petrucci2001}. Instead of this, one may also use Comptonization models such as \textsc{compTT} \citep{T94} or \textsc{nthcomp} \citep{Z96,Z99} to obtain the hot electron temperature. To use the Comptonization model to probe the corona, we replaced \textsc{cutoffpl} with the \textsc{nthcomp} in the Model-1a. The spectral model reads in \texttt{XSPEC} as (hereafter Model-1b), 

\textsc{Const1*phabs1*(zphabs2*cabs*reflect*nthcomp + Gaussian + const2*nthcomp + blackbody)}.

During the spectral fitting, we froze the seed photon temperature of \textsc{nthcomp} component at 10~eV, which is a reasonable assumption for the SMBH with $M_{\rm BH} >10^7$ \M \citep[e.g.,][]{SS1973,Makishima2000}. We verified that the variations of the seed photon temperature from 5~eV to 20~eV did not affect the spectral fitting. We obtained good fits for all the sources with Model-1b. Table~\ref{tab:pex} shows the results obtained by applying Model-1a and Model-1b in our spectral fitting.

Next, we replaced \textsc{reflect} with a torus-based physically motivated model \textsc{borus}\footnote{\url{https://sites.astro.caltech.edu/~mislavb/download/}} \citep{Balokovic2018} in Model-1a. The \borus~ model consists of a spherical homogeneous torus with two polar cutouts in a conical shape. The torus covering factor and the inclination angle are the free parameters in the model. The \borus~ model also allows us to separate the line of sight column density (\nhl) from the torus/obscuring material column density (\nht). We did not require the Gaussian function while fitting with the \borus~ model, as \borus~ self-consistently calculates the Fe~k$\alpha$ and Fe~K$\beta$ lines. In our fitting, we also fixed the torus covering factor at 0.5 and the inclination angle at 60\textdegree. The spectral model reads in \texttt{XSPEC} as (hereafter Model-2a),  

\textsc{Const1*phabs1*(zphabs2*cabs*zcutoffpl1 + borus02 + const2*cutoffpl2 + blackbody)}.

The Model-2a gave us a good fit for all the sources in our sample. We obtained \nhl, \nht, $\Gamma$ and $E_{\rm cut}$ from the spectral modelling with Model-2a. As in the case of Model-1, to  probe the corona with the Comptonized model, we replaced \textsc{cutoffpl} and \borus, with \textsc{nthcomp} and \texttt{borus12} models, respectively, in Model-2a. In the \texttt{borus12} model, the primary emission is described by \textsc{nthcomp}, replacing the \textsc{cutoffpl} model. The torus structure and geometry remain the same. This spectral model reads as (hereafter Model-2b),

\textsc{Const1*phabs1*(zphabs2*cabs*nthcomp + borus12 + const2*nthcomp + blackbody)}.

During fitting with the \borus~ model, we linked $\Gamma$, $E_{\rm cut}$ and normalization of \textsc{cutoffpl1}, \textsc{cutoffpl2}, and \borus~ together.  The spectral analysis with the \texttt{borus12} model returned with a good fit for all the sources. Table~\ref{tab:borus} shows the results obtained from our spectral fitting by applying Model-2a and Model-2b. Figure~\ref{fig:spec} shows the representative spectrum of NGC~4507 in the top panel. In the middle and bottom panels, the $\chi$ distributions are shown while using Model-1 and Model-2, respectively.

To estimate the uncertainties in the parameters, we ran the \textsc{steppar} command in {\tt XSPEC}. The uncertainties are estimated at 68\%, 90\%, and 99\% confidence levels. We quoted uncertainties at 90\% confidence level throughout the paper unless mentioned otherwise. We show confidence contours between the $\Gamma$ and the $E_{\rm cut}$ in Figure~\ref{fig:contour1} for NGC~454E, NGC~3147, NGC~3998, NGC~4102, NGC~7213, and HE~1136--2304 obtained from the fitting with Model-2a. Figure~\ref{fig:contour2} shows the confidence contours between the $\Gamma$ and the $kT_{\rm e}$, obtained from the spectral analysis of data with Model-2b for NGC~454E, NGC~3147, NGC~3998, NGC~4102, NGC~7213, and HE~1136--2304. In Figure~\ref{fig:contour1} (Figure~\ref{fig:contour2}), we selected six sources randomly to show that $E_{\rm cut}$ ($kT_{\rm e}$) could not be constrained in all sources. Detailed spectral analysis result is tabulated in Table~\ref{tab:borus}. 

We also ran Markov Chain Monte Carlo (MCMC) in \texttt{XSPEC}\footnote{\url{https://heasarc.gsfc.nasa.gov/xanadu/xspec/manual/node43.html}} to calculate the uncertainty. Using the Goodman–Weare algorithm, the chains were run with eight walkers for a total of $10^6$ steps. We discarded first 10000 steps of the chains, assuming them to be in the ``burn-in'' phase. Figure~\ref{fig:mcmc} shows the posterior distribution of the spectral parameters and errors obtained with the Model-2a and Model-2b in the left and right panels, respectively, for NGC~5033.

\subsection{Estimation of Parameters}
\label{sec:param}
The spectral analysis is carried out with four different models, with the differences in the choice of the primary continuum (\textsc{cutoffpl} or \textsc{nthcomp}), and reprocessed emission (\textsc{reflect} or \textsc{borus}). We obtained similar results with all four models. The common parameters in all four models are $\Gamma$ and \nhl. The mean value of $\Gamma$ is obtained to be $1.73\pm0.05$, $1.73\pm0.06$, $1.74\pm0.04$, and $1.75\pm0.04$, from Model-1a, Model-1b, Model-2a, and Model-2b, respectively. The mean value of \nhl~ is found to be $\log$ \nhl~ $23.39\pm0.08$, $23.38\pm0.10$, $23.40\pm0.09$, and $23.40\pm0.10$ from the spectral fitting with the Model-1a, Model-1b, Model-2a, and Model-2b, respectively. The mean values of $E_{\rm cut}$ and $kT_{\rm e}$ are also similar within uncertainties from different models. As Model-1 and Model-2 returned similar values of the spectral parameters, we used the spectral results from Model-2 in the rest of the paper or mentioned otherwise.

For 24 sources, we used black hole mass from the BAT AGN Spectroscopic Survey \citep[BASS;][]{Koss2017,Koss2022,Ricci2017bat}. For the remaining six sources, we searched for the black hole mass in the literature (see Table~\ref{tab:info}). From the spectral fitting, we estimated the intrinsic luminosity ($L_{\rm int}$) of the sources in the $2-10$~keV energy range. The bolometric luminosity is obtained by using the bolometric correction factor 10 \citep{Vasedevan2009}. We calculated the Eddington ratio as $\lambda_{\rm Edd} = L_{\rm bol}/L_{\rm Edd}$, where $L_{\rm Edd}$ is the Eddington luminosity and given by, $L_{\rm Edd} = 1.3 \times 10^{38}~(M_{\rm BH}/M_{\odot})$ \eps. 

The $\Gamma$, $kT_{\rm e}$ and $E_{\rm cut}$ were obtained from the spectral fitting. The optical depth of the corona is estimated by using the following relation \citep{Z96},

\begin{equation}
\label{eqn:tau}
\tau_{\rm e} \approx -\frac{3}{2} + \sqrt{\frac{9}{4}+\frac{3}{\theta (\Gamma - 1)(\Gamma + 2)}},
\end{equation}

where, $\theta=kT_{\rm e}/m_e c^2$ is the dimensionless temperature. 

The dimensionless compactness parameter is calculated using the following equation \citep[e.g.,][]{Fabian2015,Fabian2017},

\begin{equation}
\label{eqn:l}
l=4 \pi \frac{m_p}{m_e} \frac{R_g}{R_{\rm X}} \frac{L_{\rm X}}{L_{\rm Edd}},   
\end{equation}

where, $R_{\rm X}$ is coronal size, and $L_{\rm X}$ is the coronal luminosity in the $0.1-200$~keV energy range. The $0.1-200$~keV luminosity is calculated from the extrapolation of the best-fitted model. In this work, we used $R_{\rm X}=10~R_{\rm g}$ \citep[e.g.,][]{Fabian2015}. 

In the Compton cloud, the seed photons are up-scattered by the hot electrons and gain energy \citep[e.g.,][]{ST80,ST1985}. The mean of the energy gained by photons per scattering can be estimated by the Compton-y parameter, $y=4\theta~{\rm max}(\tau_{\rm e},\tau_{\rm e}^2)$ \citep[e.g.,][]{Rybicki1979}. On the other hand, the total energy gain also depends on the number of scattering of the photons before escaping the medium. The average number of scattering ($N_{\rm S}$) is given by, $N_{\rm S} = y/\theta$. 

A strong jet is expected in a low-accreting regime \citep{FB2004}. In general, the radio luminosity ($L_{\rm R}$) is considered a good proxy of the jet luminosity \citep[$L_{\rm jet}$; e.g.,][]{FB2004}. We collected jet luminosity ($L_{\rm jet}$ or $L_{\rm R}$ at $1.4$~GHz) of sources from the NASA Extragalactic Database (NED) archive\footnote{\url{https://ned.ipac.caltech.edu/}}.

We estimated several spectral parameters of our sample. The detailed results are presented in Table~\ref{tab:corona}.

\subsection{Correlations among coronal parameters}
\label{sec:correlation}
Correlations among several coronal parameters have been extensively studied in the past \citep[e.g.,][]{Ricci2017bat,Kamraj2018}. Here, we explored such co-dependencies among various spectral parameters. We employed Pearson, Spearman, and Kendall rank correlations to understand the relations among numerous parameters vis-\'a-vis the accretion mechanism around the LAC-AGNs. The results of our correlation study are tabulated in Table~\ref{tab:cor}. Overall, all three correlation studies yield similar results. In total, we examined 30 correlations from our study. We considered the correlation is significant if the p-value is less than $0.01$. For nine pairs of parameters, we found the corresponding p-value as $<0.01$. We found a strong anti-correlation between $kT_{\rm e}$ and $E_{\rm cut}$. The $N_{\rm S}$ is observed to be strongly correlated with $kT_{\rm e}$ and $E_{\rm cut}$. We found moderate anti-correlations and correlations for three pairs of parameters each.

\section{Results and Discussion}
\label{sec:res}
We studied a sample of AGNs with low Eddington ratio ($\lambda_{\rm Edd}<10^{-3}$) to understand their coronal properties at low accretion regime. In our study, we used combined \xmm, \swift, and \nustar~ spectra in the $0.5-150$~keV energy range. From the spectral study, we obtained diverse spectral parameters and correlations among them. 

\subsection{Constraints on the Coronal Parameters}
\label{sec:cor-par}
The corona is characterized by several parameters, namely the photon index ($\Gamma$), hot electron temperature ($kT_{\rm e}$), cutoff energy ($E_{\rm cut}$), and optical depth ($\tau_{\rm e}$). We obtained $\Gamma$, $kT_{\rm e}$ and $E_{\rm cut}$ from the spectral analysis, while $\tau_{\rm e}$ is obtained using Equation~\ref{eqn:tau}. Figure~\ref{fig:hist2} shows the distribution of $E_{\rm cut}$, $kT_{\rm e}$ and $\tau_{\rm e}$ in our sample. 

We are able to constrain $E_{\rm cut}$ in 22 sources out of a total of 30 sources considered in our sample. The $E_{\rm cut}$ is distributed in a wide range of $\sim 50-500$~keV in our sample. The broad parameter space of $E_{\rm cut}$ is consistent with the other recent studies \citep[e.g.,][]{Ricci2018,Balokovic2020,Hinkle2021}. We found that the lower limit of $E_{\rm cut}$ is below 50~keV for two sources when $E_{\rm cut}$ is not constrained. The median value of $E_{\rm cut}$ for our sample is found to be  $238\pm93 $~keV with mean $<E_{\rm cut}> = 241 \pm 84$~keV, when $E_{\rm cut}$ is constrained. However, these values do not represent the whole sample, as the sources with the unconstrained $E_{\rm cut}$ are not considered.

To constrain the mean and median of the whole sample, we performed 1000 Monte Carlo simulations for each value of $E_{\rm cut}$. For each simulation, the values of $E_{\rm cut}$ are substituted with the values selected randomly from a Gaussian distribution with the standard deviation given by the uncertainty. The lower limits (L) are substituted with the values randomly selected from a uniform distribution in the interval of [L, $E_{\rm C, max}$], where $E_{\rm C, max}=1000$~keV. For each run, we calculated the median of all values and used the mean of 1000 simulations \citep[see, ][for details]{Ricci2017bat,Ricci2018}. We find that the mean of $E_{\rm cut}$ is $284\pm 102$~KeV, while the median is $267\pm 110$~keV. Our results are consistent with the results of \citet{Ricci2017bat,Ricci2018} and the studies of the cosmic X-ray background, which suggest that the mean cutoff energy of AGNs should be $\lesssim 300$~keV \citep[e.g.,][]{Gilli2007,Ueda2014,Ananna2020ApJ}.  

Using \textsc{Model-2}, we additionally constrained $kT_{\rm e}$ for those 22 sources. We found a lower limit for the other eight sources. Analogous to $E_{\rm cut}$, $kT_{\rm e}$ was also obtained in a broad range between $\sim 10-300$~keV which was found in other studies \citep[e.g.,][]{Tortosa2018,Akylas2021}. We obtained the lower limit of $kT_{\rm e}$ as $15$~keV for IC~4518A, which is the lowest value among the sources in our samples. We calculated the mean and median of $kT_{\rm e}$ by running 1000 Monte Carlo simulations, as mentioned in the previous paragraph. We considered the maximum value of $kT_{\rm e}$ as 500~keV for the sources with the lower limit in the simulation. We found the mean value of our samples as $<kT_{\rm e}> = 126 \pm 54$~keV with a median at $110 \pm 45$~keV.

As $\tau_{\rm e}$ is calculated using $kT_{\rm e}$, we also derived upper limits on $\tau_{\rm e}$ for six observations. Excluding the upper limits, $\tau_{\rm e}$ is observed to vary within the range of $\sim 0.5-3$. The mean of the optical depth is estimated to be $<\tau_{\rm e}> = 1.77\pm 0.76$ with median $\tau_{\rm e} = 1.47 \pm 0.58$ in our sample. 

The corona remains hot for low mass accretion rate AGNs as the cooling is inefficient. The hot corona also leads to high $E_{\rm cut}$ and optically thin medium ($\tau_{\rm e} <1$). \citet{Ricci2018} found the median of cutoff energy, $E_{\rm cut} = 160 \pm 41$~keV for $\lambda_{\rm Edd} > 0.1$, and $E_{\rm cut} = 370 \pm 51$~keV for $L/L_{\rm Edd} < 0.1$. As we explored an even lower Eddington ratio ($\lambda_{\rm Edd} < 0.001$) regime, the median of the cutoff energy is expected to be higher. However, this was not observed for our sample of low Eddington ratio AGN.


\subsection{Dependence of the coronal properties on the Eddington ratio}
\label{sec:lambda}
In the current work, we studied a sample of AGNs with low $\lambda_{\rm Edd}$ to understand the coronal properties in the low accretion regime. The $\Gamma-\lambda_{\rm Edd}$ relation has been studied widely in the past \citep[e.g.,][]{Gu2009,Yang2015}. A positive correlation is observed in HAC-AGN \citep[e.g.,][]{Brightman2013,Risaliti2009,Shemmer2006,Shemmer2008,AJ2020,Jana2021} while a negative correlation is found in the LLAGNs \citep[e.g.,][]{Gu2009,Younes2011,Hernandez2013}. The accretion mechanism differs in the low accretion state from the high accretion state. The opposite correlation indicates that the accretion mechanisms in different luminosity states are distinct. 

The thin disc-corona model naturally explains the positive correlation in the high accretion state \citep[e.g.,][]{Yang2015}. In the high accreting regime ($\lambda_{\rm Edd}>10^{-3}$), as the accretion rate increases, the number of seed photons increases, which cools the corona efficiently, producing the soft spectra. Contrary to that, the negative correlation in the low accretion state ($\lambda_{\rm Edd}<10^{-3}$) could be explained with a hybrid truncated thin disc associated with hot accretion flow/corona and jet models \citep[e.g.,][]{Gardner2013,Qiao2013,Yang2015}. In this scenario, due to the lack of matter supply, the inner disc evaporates into a hot accretion flow or corona \citep[e.g.,][]{Esin1997,Yuan2014,Yang2015}. As the mass accretion rate (or $\lambda_{\rm Edd}$) increases, the electron density and magnetic field strength increase, which in turn increases the synchrotron self-absorption depth. The self-absorbed synchrotron emission provides the seed photons for Comptonization. In this case, the hard X-ray flux ($L_{\rm X}$) increases more rapidly than the seed photon flux ($L_{\rm seed}$), which implies a negative correlation of $L_{\rm seed}/L_{\rm X}$ with $L_{\rm X}$. This leads to the negative correlation of the $\Gamma$ and $L_{X}$ or $\lambda_{\rm Edd}$ \citep[e.g.,][]{Yang2015}. If the accretion rate further reduces ($\lambda_{\rm Edd}<10^{-6.5}$), the synchrotron emission from the jet dominates, leading to a saturation of the photon index at $\Gamma \sim 2$ \citep[e.g.,][]{Plotkin2013,Yang2015}.

In our sample, $\lambda_{\rm Edd}$ spans the range $10^{-6.5}<\lambda_{\rm Edd}<10^{-3}$, where a negative correlation of $\Gamma-\lambda_{\rm Edd}$ is expected. However, we did not find a significant correlation between $\Gamma$ and $\lambda_{\rm Edd}$. The Pearson correlation coefficient between $\Gamma-\lambda_{\rm Edd}$ is $-0.22$ with p-value 0.24. Figure~\ref{fig:cor-edd} shows the variation of $\Gamma$ with the $\lambda_{\rm Edd}$. The solid red line represents the best linear fit. Using the linear regression method, we obtained $\Gamma = (-0.04\pm 0.03)\log \lambda_{\rm Edd} + (1.59\pm 0.13)$. The observed relation is weaker than the one found in previous studies. For examples, \citet{Gu2009} found $\Gamma=(-0.09\pm0.03) \log \lambda_{\rm Edd}+(1.55\pm 0.07)$, \citet{Younes2011} observed $\Gamma=(-0.31\pm0.06) \log \lambda_{\rm Edd}+(0.11\pm 0.40)$, \citet{Jang2014} pointed $\Gamma = (-0.18\pm0.04) \log \lambda_{\rm Edd}+(0.66\pm 0.25)$, and \citet{She2018} obtained $\Gamma=(-0.15\pm0.05) \log \lambda_{\rm Edd}+(1.0\pm 0.03)$. Most of the previous studies were conducted using \chandra~ or \xmm~ observations, which have a limited band-pass. \citet{Trakhtenbrot2017} found a shallower slope of the $\Gamma-\lambda_{\rm Edd}$ correlation with respect to previous studies, when considering the results obtained by broad-band X-ray spectroscopy. In the current work, we used high-quality broad-band spectra, though our sample is limited to 30 sources. Thus, we are unable to make a firm conclusion on the correlation/anti-correlation of $\lambda_{\rm Edd}-\Gamma$. Recently, \citet{Diaz2023} studied a sample of LLAGNs and found similar results. They argued that due to the small number of sources, they did not find a statistically significant correlation of $\Gamma-\lambda_{\rm Edd}$. We also inspected whether other spectral parameters are correlated with the $\lambda_{\rm Edd}$ and did not observe any correlation between $\lambda_{\rm Edd}$ and $kT_{\rm e}$, $\tau_{\rm e}$ or $E_{\rm cut}$.

\subsection{Dependency of the Coronal Properties on the $kT_{\rm e}$}
\label{sec:kt}
We found that $kT_{\rm e}$ and $E_{\rm cut}$ are strongly correlated (see Table~\ref{tab:cor}) in our sample. The linear fitting yields $E_{\rm cut}=(2.10\pm0.12) kT_{\rm e} + (29.4\pm12.1)$. Figure~\ref{fig:cor-kt} shows that all the 22 sources with constrained $E_{\rm cut}$ and $kT_{\rm e}$ lie within the $E_{\rm cut} = 2 kT_{\rm e}$ and $E_{\rm cut} = 3 kT_{\rm e}$ lines. This agrees with the empirical approximation of $E_{\rm cut} \approx 2-3~kT_{\rm e}$ \citep{Petrucci2001}. However, \citet{Middei2019} argued that the empirical relation only holds for low $\tau_{\rm e}$ and low $kT_{\rm e}$. They suggested that if the origin of X-rays is other than the thermal Comptonization, for example, synchrotron self Comptonization, the relation $E_{\rm cut} \sim 2-3~kT_{\rm e}$ may not hold. The deviation from this relation has been observed in a few sources \citep[e.g.,][]{Pal2022}. We tested this relation from the spectral modelling with the \textsc{reflect} model (Model-1a \& Model-1b). We obtained $E_{\rm cut}=(2.18\pm0.16)kT_{\rm e} + (25.\pm17.2)$, which is similar to the findings with the \textsc{borus} model (Model-2a \& Model-2b). Nonetheless, our study found $E_{\rm cut} \approx 2~kT_{\rm e}$ is an acceptable approximation for the LAC-AGNs.

We obtained a weak positive correlation between $kT_{\rm e}$ and $\Gamma$, with the Pearson correlation coefficient of $0.34$ with a p-value of $0.01$. For purely thermal Comptonization, a negative correlation between $kT_{\rm e}$ and $\Gamma$ is expected. However, if there are non-thermal seed photons, e.g., synchrotron emission from a jet, a negative correlation may not hold \citep[e.g.,][]{Yang2015}. Nonetheless, the observed relation indicates a complex process for the X-ray emission other than thermal Comptonization.

We calculated the average number of scatterings the photons suffered before escaping the Compton cloud (see \S~\ref{sec:param}). We find that $N_{\rm s}$ is anti-correlated with $\Gamma$ having a Pearson correlation coefficient of $-0.65$ with $p<0.01$, as presented in the earlier works \citep[e.g.,][]{ST80,PSS1983}. $N_{\rm S}$ is also found to be strongly anti-correlated with $kT_{\rm e}$ and $E_{\rm cut}$. The Pearson correlation index is obtained to be $-0.85$ with the p-value of $<0.01$ for $kT_{\rm e}$, and $-0.85$ with the p-value of $<0.01$ for $E_{\rm cut}$, respectively. This is expected as a high $kT_{\rm e}$ would lead the corona to be optically thin, hence, the lower value of $N_{\rm S}$. The anti-correlation between $N_{\rm S}$ and $kT_{\rm e}$ is also consistent with the previous simulations \citep[e.g.,][]{Chatterjee2017a, Chatterjee2017b}. Compton scattering could induce X-ray polarization. The variation in the polarization caused by repeated scatterings could be detected above the minimum detectable polarization with the ongoing \textit{Imaging X-ray Polarimetry Explorer} \citep{Weisskopf2016} or future X-ray polarimetry missions, such as \textit{XPoSat}.

\begin{figure}
\centering
\includegraphics[width=8.5cm]{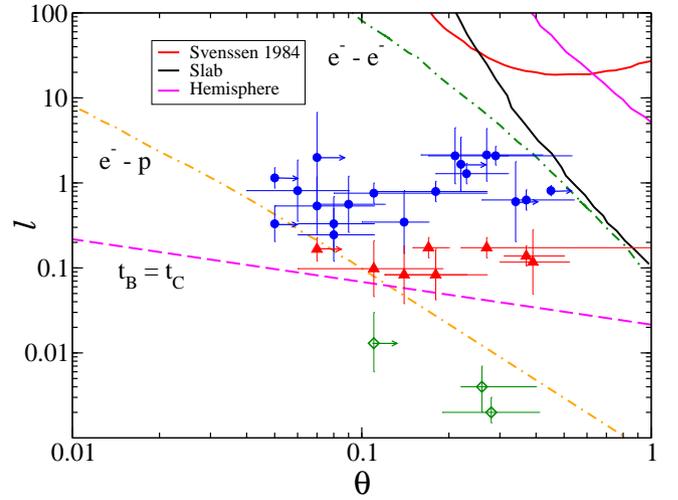}
\caption{Compactness-temperature ($l-\theta$) diagram for our sample. Solid black and red lines correspond to pair production lines for slab and hemispherical geometries from \citet{Stern1995}, respectively. The solid magenta line represents the pair production line from \citet{Svensson1984}. The orange dash-dot, green dashed-dot, and magenta dashed lines represent the region where electron-electron coupling, electron-proton coupling, and bremsstrahlung cooling dominate, respectively. The blue circles, red triangles, and green diamonds represent the AGNs with $-3>\log \lambda_{\rm Edd}>-4$, $-4>\log \lambda_{\rm Edd}>-5$, and $\log \lambda_{\rm Edd}<-5$, respectively. The arrows represent the lower limit.}
\label{fig:l-theta}
\end{figure}

\subsection{The $l-\theta$ Plane}
\label{sec:l-theta}
We constructed the compactness-temperature ($l-\theta$) diagram in Figure~\ref{fig:l-theta}. Solid black and red lines correspond to the pair production lines for slab and hemispherical geometries from \citet{Stern1995}. The solid magenta line represents the pair production line from \citet{Svensson1984}. The compactness ($l$) is calculated using Equation~\ref{eqn:l}. 

The pair production is thought to be a fundamental process in AGN coronae due to photon-photon collisions \citep[e.g.,][]{Svensson1982a,Svensson1982b,Guilbert1983}. This process could, in fact, lead to a runway pair production, which might act as a thermostat for the corona \citep[e.g.,][]{Bisnovatyi1971,Svensson1984,Zdziarski1985,Fabian2015,Fabian2017}. If that were the case, then the AGN would be expected to lie below the pair line. We found that all the sources are located below the theoretical pair lines for slab and hemispherical geometry. The sources in our sample are located around the electron-electron ($e^--e^-$) and electron-proton ($e^--p$) coupling lines, which could indicate the processes responsible for the cooling. We also found that three sources lie below the bremsstrahlung cooling line ($t_{\rm B}=t_{\rm C}$). All three sources, namely NGC~3718, NGC~3998, and UGC~12282, have $\lambda_{\rm Edd} < 10^{-5}$ and, as the Eddington ratio is directly proportional to the compactness, which leads to the low compactness \citep{Ricci2018}. 

\begin{figure}
\centering
\includegraphics[width=8.5cm]{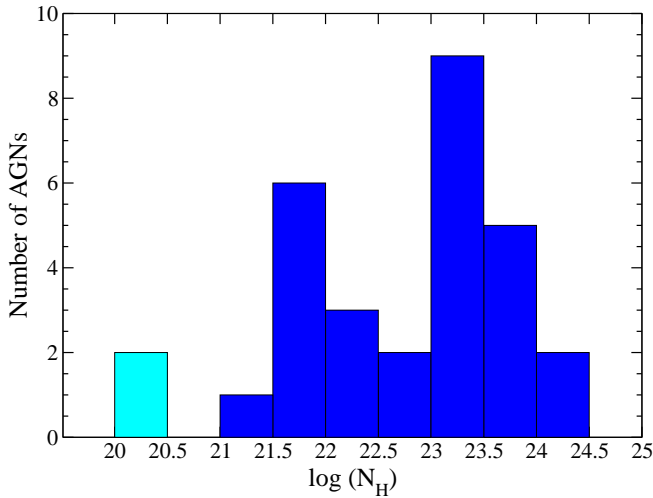}
\caption{Histogram of the line of sight column density of our sample. The blue and cyan colours represent the constrained value and upper limit of $N_{\rm H}$, respectively.}
\label{fig:nh-hist}
\end{figure}

\subsection{Obscuration Properties}
\label{sec:torus}
The covering factor of the circumnuclear obscuring materials has been found to decrease with increasing accretion rates \citep[e.g.,][]{Ueda2003,Treister2008}. However, recent work has shown that the obscuring material in nearby AGN is regulated by the Eddington ratio \citep{Ricci2017nat}. It has been found that the covering factor is $\sim 85\%$ for $\lambda_{\rm Edd} \sim 10^{-4}-10^{-1.5}$, and sharply decreases at $\lambda_{\rm Edd}>10^{-1.5}$ \citep{Ricci2017nat}. The obscuring material is expected to disappear at very low accretion rates \citep[e.g.,][]{Elitzur2008} due to the lack of outflowing material \citep[e.g.,][]{Elitzur2008}. \cite{Ricci:2022} suggested that an inactive AGN ($\lambda_{\rm Edd}<<10^{-4}$) starts accreting following an inflow of gas and dust (see also \citealp{Ricci2017nat}). This increases both $N_{\rm H}$ and $\lambda_{\rm Edd}$. When $\lambda_{\rm Edd}$ reaches a critical value ($\lambda_{\rm Edd} \sim 10^{-1.5}$), the radiation pressure blows away the obscuring material. The AGN spends some time as unobscured ($N_{\rm H}<10^{22}$ \pcm) before moving back to the low $\lambda_{\rm Edd}$ state with low $N_{\rm H}$.

Figure~\ref{fig:nh-hist} shows the histogram of our sample's line of sight column density. In our sample, we found that nine sources are unobscured (\nhl $<10^{22}$ \pcm) and 21 sources are obscured (\nhl $>10^{22}$ \pcm). Among the {21} obscured sources, two sources are Compton-thick (\nhl $>10^{24}$ \pcm) and 19 sources are Compton-thin (\nhl $=10^{22-24}$ \pcm). We found two unobscured sources and one obscured source (one of them in CT) in an extremely low accretion region ($\lambda_{\rm Edd}<10^{-5}$). On the other hand, at $10^{-5}<\lambda_{\rm Edd}<10^{-3}$, we observed that seven sources are unobscured, and 20 sources are obscured. If we move towards the low accretion region ($\lambda_{\rm Edd}<-5$), the fraction of obscured sources drops from $\sim 73^{+8}_{-9}$\% to $\sim 39^{+23}_{-20}\%$\footnote{The fractions are computed following \citet{Cameron2011}, and the reported uncertainties represent the 16th and 84th quantiles of a binomial distribution.}. The increasing fraction of unobscured sources towards the low-accretion regime supports the Eddington ratio regulated unification model \citep[e.g.,][]{Ricci2017nat,Ricci:2022}.

\begin{figure}
\centering
\includegraphics[width=8.5cm]{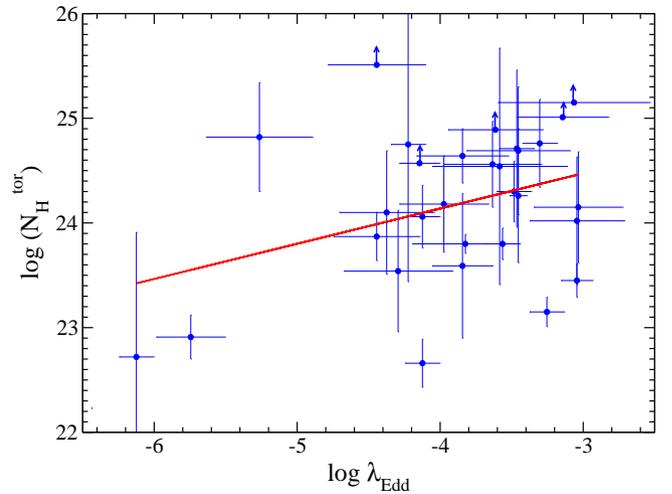}
\caption{The variation of average density of the obscuring material ($N_{\rm H}^{\rm tor}$) as a function of the Eddington ratio ($\lambda_{\rm Edd}$). The red line represent the linear best-fit.}
\label{fig:edd-nh}
\end{figure}

We obtained the average density of the obscuring material ($N_{\rm H}^{\rm tor}$), which is responsible for the reprocessing emission. The median of $N_{\rm H}^{\rm tor}$ is found to be $\log N_{\rm H}^{\rm tor}=24.22 \pm 0.45$, which is higher than the $N_{\rm H}^{\rm los}$. The median of the line of sight column density is $\log N_{\rm H}^{\rm los} = 23.02\pm 0.08$. We plot the variation of $N_{\rm H}^{\rm tor}$ as a function of $\lambda_{\rm Edd}$ in Figure~\ref{fig:edd-nh}. The linear regression analysis found that $\log N_{\rm H}^{\rm tor}=(0.34\pm0.04) \log{\lambda_{\rm Edd}}+(25.5\pm0.5)$, suggesting a positive relation between  $\lambda_{\rm Edd}$ and $N_{\rm H}^{\rm tor}$. This suggests that at low Eddington ratio, the average column density decreases. \citet{Diaz2023} also found a similar relation between $N_{\rm H}^{\rm tor}$ and $\lambda_{\rm Edd}$. The relation between $N_{\rm H}^{\rm tor}$ and $\lambda_{\rm Edd}$ also supports the Eddington ratio regulated unification and growth model \citep[e.g.,][]{Ricci:2022}.

\subsection{Reprocessed Emission}
We obtained the reflection parameter ($R$) from Model-1a and Model-1b. From our analysis, it is found that $R$ could be constrained only in 7 sources out of a total of 30 sources. As in case of $E_{\rm cut}$ and $kT_{\rm e}$, we calculated the mean of $R$ by running 1000 Monte Carlo simulations, with the range of $R$ between 0 to 10. We found the mean $<R>=0.25\pm0.08$ with a median of $0.26\pm 0.09$. \citet{Ricci2017bat} found a higher value of median of $R$ as $0.53\pm0.09$ with a sample of 838 BAT AGNs. Our finding is consistent with the fact that the low-accreting AGNs show weak reflection \citep[e.g.,][]{Younes2011,Ptak2004}.

\begin{figure}
\centering
\includegraphics[angle=270,width=9.0cm]{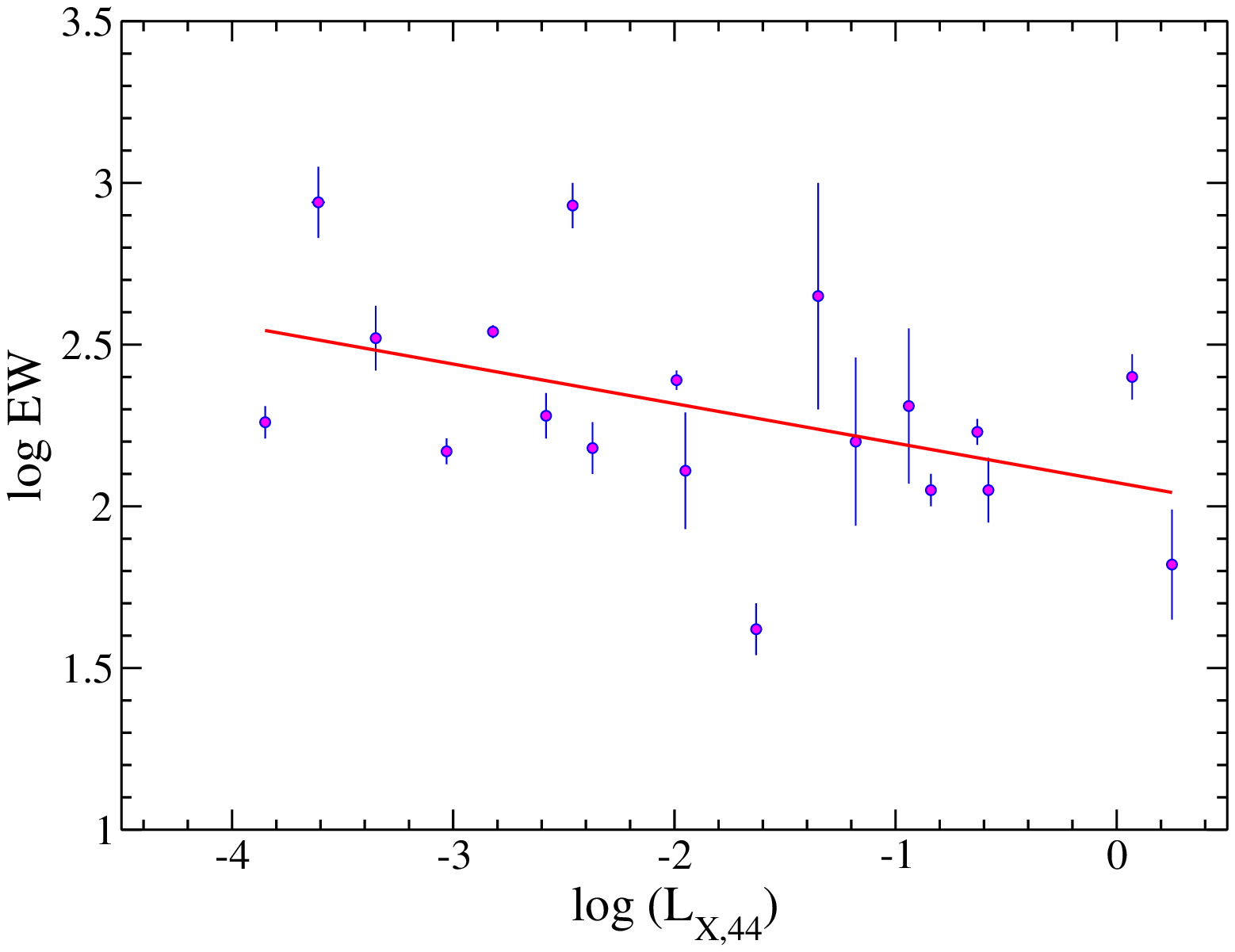}
\caption{X-ray Baldwin effect. The EW of iron K$\alpha$ line is plotted as a function of $2-10$ keV X-ray luminosity ($L_{\rm X, 44}$). The EW and $L_{X,44}$ are in the unit of eV, and $10^{44}$ \eps, respectively. The red line represent the linear best-fit, with $\log {\rm EW} = (-0.12\pm0.09)\log L_{\rm X, 44} + (2.1 \pm 0.1).$}
\label{fig:xbw}
\end{figure}

We modelled the Fe K-emission line with a Gaussian function while fitting the data with Model-1. Out of total of 30 sources, a Gaussian line is required in 28 objects. We did not find the iron K$\alpha$ line for two sources: NGC~3147 and NGC~3998. We could constrain the equivalent width (EW) for 19 sources. We tested the so-called `X-ray Baldwin effect', i.e., the correlation of EW with the X-ray luminosity in our sample \citep{Iwasawa1993}. Figure~\ref{fig:xbw} shows the EW of Fe K$\alpha$ line as a function of X-ray luminosity ($L_{\rm X, 44}$). The linear regression analysis returned as $\log {\rm EW} = (-0.12\pm0.09)\log L_{\rm X, 44} + (2.1 \pm 0.1)$, where EW and $L_{X,44}$ are in the unit of eV, and $10^{44}$ \eps, respectively. This relation is consistent with the previous studies of the X-ray Baldwin effect \citep[e.g.,][]{Bianchi2007,Ricci2013}. We also checked the relation of EW with $\lambda_{\rm Edd}$. The linear best-fit result returned with $\log {\rm EW} = (-0.15\pm0.10)\log \lambda_{\rm Edd} + (1.74 \pm 0.52).$ Our result is consistent with the previous studies \citep[e.g.,][]{Ricci2013}. The observed relation suggests that the reprocessing mechanism of LAC-AGNs is similar to the HAC-AGNs.

We did not find any correlation or anti-correlation of EW with the $L_{\rm 2-10}$~keV, $L_{\rm X, 44}$, and $\lambda_{\rm Edd}$. The small sample size could be the reason for not observing any correlation/anti-correlation of EW with other parameters.

\begin{figure}
\centering
\includegraphics[width=8.5cm]{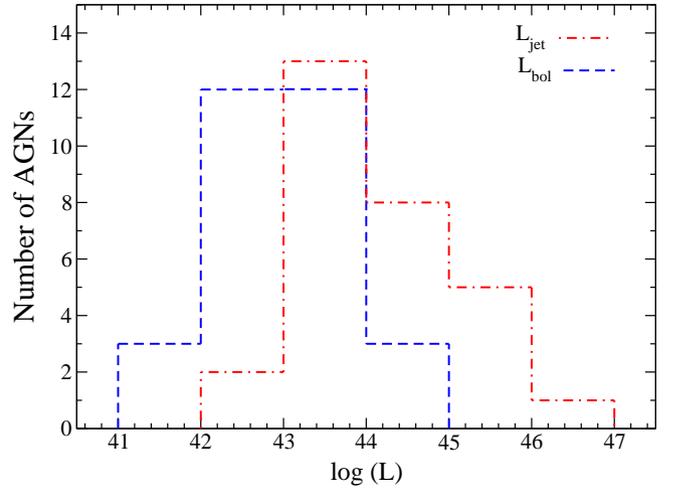}
\caption{Histogram of the observed jet luminosity ($L_{\rm jet}$) and bolometric luminosity ($L_{\rm bol}$). The red dashed and solid blue lines represent the $L_{\rm jet}$ and $L_{\rm bol}$, respectively.}
\label{fig:jet-hist}
\end{figure}

\label{sec:jet}
\subsection{Jet}
In LAC-AGNs X-ray radiation is believed to be produced in a radiatively inefficient flow \citep[e.g.,][]{Naayan1994,Quataert2001,Nemen2014}, or from the base of the jet \citep[e.g.,][]{Markoff2001,Falcke2004}. The observed jet luminosity ($L_{\rm jet}$)\footnote{$L_{\rm jet}$ is calculated from the $L_{\rm R}$.} is tabulated in Table~\ref{tab:cor}. Figure~\ref{fig:jet-hist} shows the histogram plots for the $L_{\rm bol}$ and the $L_{\rm jet}$ of our sample. The red dashed, and solid blue lines represent the $L_{\rm jet}$ and $L_{\rm bol}$, respectively. We observed that the $L_{\rm jet}$ is higher than the $L_{\rm bol}$\footnote{The $L_{\rm bol}$ is calculated from the $2-10$~keV X-ray luminosity (see \S~\ref{sec:analysis})} for every source in our sample. The $L_{\rm jet}$ is found to be $\sim 0-3$ orders of magnitude higher than the $L_{\rm bol}$. If we consider $\sim 5\%$ power of the jet is radiated away \citep[e.g.,][]{Blandford1979,Fender2001}, the total jet power ($Q_{\rm jet}$) would be $\sim 0-4$ orders of magnitude higher than the bolometric luminosity ($L_{\rm bol}$). This is not uncommon for LLAGNs: \citet{Nagar2005} studied a sample of LLAGNs and found $Q_{\rm jet}$ (also $L_{\rm jet}$) exceeds the $L_{\rm bol}$ by $0-4.5$ magnitude with a mean around three orders in their sample. This study suggests that the jet luminosity greatly surpasses the accretion power in the LAC-AGNs.

\begin{figure}
\centering
\includegraphics[width=8.5cm]{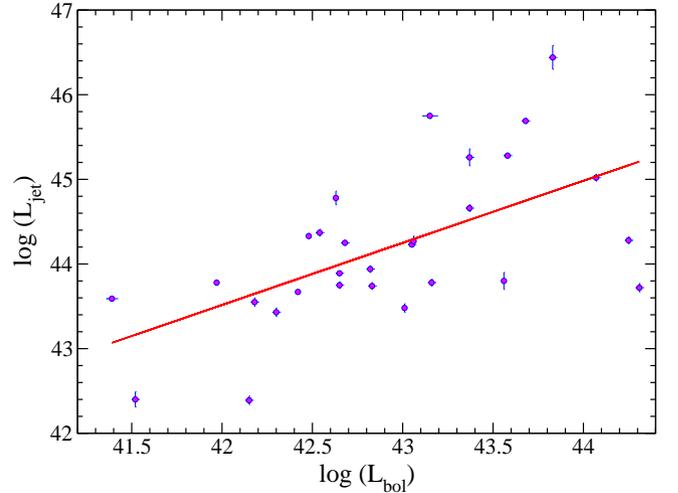}
\caption{Variation of the jet luminosity $L_{\rm jet}$ as a function of Eddington ratio ($\lambda_{\rm Edd}$). The solid red line represents the best fit with linear regression. The slope of the fit is $0.68\pm0.12$.}
\label{fig:jet}
\end{figure}

Figure~\ref{fig:jet} shows the variation of the jet luminosity ($L_{\rm jet}$) with the bolometric luminosity ($L_{\rm bol}$). The solid red line represents the best linear fit $\log (L_{\rm jet})=(0.68\pm0.12)\log(L_{\rm bol})+(12.7\pm8.1)$ from the linear regression analysis. This relation is consistent with the standard radio-X-ray correlation of coefficient of $\sim 0.6-0.7$ \citep[e.g.,][]{Corbel2000,Corbel2003,Merloni2003,Gallo2003,Gallo2014,Kording2006}. The standard correlation is the relation between the radio and X-ray flux in the low hard state of black holes. Thus, the similar relation of $L_{\rm jet}-L_{\rm bol}$ for the present sample indicates a radiatively inefficient accretion flow in the LAC-AGNs.

\begin{table*}
\caption{Correlation among different parameters.}
\label{tab:cor}
\begin{tabular}{cccccccccc}
\hline
 & & Pearson Correlation & & Spearman Correlation & & Kendall Correlation  \\
Parameter-1 & Parameter-2 & $\rho$ & p & R & p & $\tau$ & p \\
\hline
$\lambda_{\rm Edd}$ & $\Gamma$ & --0.22 & 0.24 & --0.16 & 0.40 & --0.12 & 0.36 \\
$\lambda_{\rm Edd}$ & $E_{\rm cut}$ &  --0.02 & 0.90 & 0.01 & 0.99 & --0.01 & 0.94 \\
$\lambda_{\rm Edd}$ & $kT_{\rm e}$ & --0.07 & 0.71 & --0.01 & 0.95 & --0.02 & 0.89\\
$\lambda_{\rm Edd}$ & $\tau_{\rm e}$ & 0.18 & 0.33 & --0.02 & 0.90 & --0.03 & 0.84\\
$\lambda_{\rm Edd}$ & $M_{\rm BH}$ & --0.45 & <0.01 & --0.46 & 0.01 & --0.32 & 0.02\\
$kT_{\rm e}$ & $\Gamma$ & 0.34 & 0.01 & 0.40 & 0.02 & 0.29 & 0.02 \\
$kT_{\rm e}$ & $E_{\rm cut}$ & 0.97 & <0.01 & 0.95 & <0.01 & 0.85 & <0.01 \\
$\Gamma$ & $E_{\rm cut}$ & 0.32 & 0.08 & 0.40 & 0.03 & 0.28 & 0.03 \\
$\Gamma$ & $\tau_{\rm e}$ & --0.64 & <0.01 & --0.56 & <0.01 & --0.41 & <0.01 \\
$M_{\rm BH}$ & $\tau_{\rm e}$ & --0.11 & 0.61 & --0.04 & 0.83 & 0.01 & 0.93 \\
$M_{\rm BH}$ & $kT_{\rm e}$ & 0.04 & 0.84 & 0.03 & 0.87 & 0.03 & 0.86 \\
$M_{\rm BH}$ & $E_{\rm cut}$ & --0.07 & 0.69 & --0.02 & 0.92 & --0.02 & 0.91\\
$L_{\rm bol}$ & $M_{\rm BH}$ & 0.49 & <0.01 & 0.51 & <0.01 & 0.40& <0.01 \\
$L_{\rm bol}$ & $\Gamma$ & --0.21 & 0.26 & --0.23 & 0.22 & --0.017 & 0.18 \\
$L_{\rm bol}$ & $kT_{\rm e}$ & --0.11 & 0.56 & --0.07 & 0.70 & --0.06 & 0.62 \\
$L_{\rm bol}$ & $E_{\rm cut}$ & --0.10 & 0.59 & --0.07 & 0.71 & --0.06 & 0.68 \\
$L_{\rm bol}$ & $\tau_{\rm e}$ & 0.09 & 0.64 & 0.13 & 0.49 & 0.09 & 0.49 \\
$L_{\rm bol}$ & $\lambda_{\rm Edd}$ & 0.46 & <0.01 & 0.41 & <0.01 & 0.29 & 0.01 \\
$\lambda_{\rm Edd}$ & $N_{\rm H}$ & 0.08 & 0.68 & --0.12 & 0.52 & --0.09 & 0.49 \\
$L_{\rm bol}$ & $N_{\rm H}$ & 0.06 & 0.76 & 0.02 & 0.90 & 0.01 & 0.99 \\
$N_{\rm S}$ & $\Gamma$ & --0.65 & <0.01 & --0.56 & <0.01 & --0.41& <0.01 \\
$N_{\rm S}$ & $kT_{\rm e}$ & --0.85 & <0.01 & -0.97 & <0.01 & --0.89 & <0.01 \\
$N_{\rm S}$ & $E_{\rm cut}$& --0.84 & <0.01 & --0.91 & <0.01 & -0.76  & <0.01 \\
$N_{\rm S}$ & $\lambda$ & 0.19 & 0.31 & --0.02 & 0.90 & --0.03 & 0.84 \\
$N_{\rm S}$ & $L_{\rm bol}$ & 0.11 & 0.57 & 0.13 & 0.49 & 0.09 & 0.50 \\
$N_{\rm S}$ & $M_{\rm BH}$ & --0.08 & 0.66 & --0.04 & 0.83 & 0.01 & 0.93 \\
$L_{\rm bol}$&$L_{\rm jet}$ & 0.60 & <0.01 & 0.61 & <0.01 & 0.49 & <0.01 \\
$EW$ & $L_{\rm bol}$& --0.58 & 0.01 & --0.52 & 0.02 & --0.39 & 0.02 \\
$EW$ & $\lambda_{\rm Edd}$ & --0.25 & 0.29 & --0.14 & 0.54 & --0.11 & 0.55 \\
$EW$ & $L_{\rm X, 44}$ & --0.45 & 0.06 & --0.31 & 0.06 & --0.44 & 0.05 \\
\hline
\end{tabular}
\end{table*}

\section{Conclusion and Summary}
\label{sec:summary}
We studied 30 low-accreting AGNs ($\lambda_{\rm Edd} < 10^{-3}$) using combined \swift, \xmm, and \nustar~ data in $0.5-150$~keV range. For the spectral analysis, we used the convolution model \textsc{reflect} and torus-based physically motivated \textsc{borus} model combined with either the \textsc{cutoffpl} or the \textsc{nthcomp} model for the continuum. Several parameters, namely the photon index, cutoff energy, and hot electron temperature of the corona, are estimated directly from the spectral fitting. Other parameters, such as the optical depth, the number of scatterings, and compactness, are calculated using spectral parameters. We inspected correlations among several parameters to understand the accretion dynamics in the low accreting region. 

We summarize our work as follows.

\begin{enumerate}
\item We studied 30 low-accreting AGNs ($\lambda_{\rm Edd} < 10^{-3}$) using combined \swift, \xmm, and \nustar~ data to understand the accretion properties at low accretion region.
\item We did not find any significant correlation between the photon index ($\Gamma$) and Eddington ratio ($\lambda_{\rm Edd}$), contrary to the previous studies in the low-accretion domain. 
\item We found that the hot electron temperature ($kT_{\rm e}$) is related to the cutoff energy ($E_{\rm cut}$) as $E_{\rm cut}=(2.10\pm0.12)kT_{\rm e}+(29.4\pm12.1)$.
\item We noticed that all the sources are located well below the pair production line in the compactness-temperature ($l-\theta$) diagram. We note that the cooling process is complex in the low accretion region.
\item We observed the so-called `X-ray Baldwin effect' in low-accretion regime. The EW of the Fe K$\alpha$ line is found to be related with the X-ray luminosity ($L_{\rm X, 44}$) and Eddington ratio ($\lambda_{\rm Edd}$) as $\log {\rm EW} = (-0.12\pm0.09)\log L_{\rm X, 44} + (2.1 \pm 0.1)$ and $\log {\rm EW} = (-0.15\pm0.10)\log \lambda_{\rm Edd} + (1.74 \pm 0.52)$.
\item The jet luminosity ($L_{\rm jet}$) is related with the bolometric luminosity as $L_{\rm jet} \propto L_{\rm bol}^{0.7}$. This relation is consistent with the standard radio-X-ray correlation for Galactic black hole X-ray binary in the Low hard state. This supports the presence of a radiatively inefficient accretion flow in the LAC-AGNs.
\item We observed that the fraction of the unobscured sources increases as the Eddington ratio decreases. This support the Eddington ratio regulated unification model of AGNs.
\end{enumerate}

In this work, we studied the coronal properties of AGNs with the Eddington ratio ranges in $\lambda_{\rm Edd} \sim 10^{-6.5} - 10^{-3}$. In the future, we will add more AGNs to our sample with an even lower Eddington ratio. Future broadband hard X-ray missions, such as \textit{HEX-P} \citep{Madsen2018}, could allow us to constrain $E_{\rm cut}$ with better accuracy and expand our understanding of such systems profoundly. On the other hand, the large effective area and high throughput of \textit{Colibr\`i} \citep{Heyl2019, Caiazzo2019} would be able to extend the population of LLAGNs as well as provide crucial information related to the line of sight $N_H$ distribution \citep{Ricci2017nat} of them.

\section*{Acknowledgements}
\label{sec:ack}
We thank the anonymous reviewers for their constructive suggestions which helped to improve the paper. AJ and HK acknowledge the support of the grant from the Ministry of Science and Technology of Taiwan with the grand number MOST 110-2811-M-007-500 and MOST 111-2811-M-007-002. HK acknowledge the support of the grant from the Ministry of Science and Technology of Taiwan with the grand number MOST 110-2112-M-007-020 and MOST-111-2112-M-007-019. AC and SSH are supported by the Canadian Space Agency and the Natural Sciences and Engineering Research Council of Canada. CR acknowledges support from the Fondecyt Iniciacion grant 11190831 and ANID BASAL project FB210003. Research at Physical Research Laboratory is supported by the Department of Space, Government of India, for this work.  This research has made use of data and/or software provided by the High Energy Astrophysics Science Archive Research Center (HEASARC), which is a service of the Astrophysics Science Division at NASA/GSFC and the High Energy Astrophysics Division of the Smithsonian Astrophysical Observatory. This work has made use of data obtained from the {\it NuSTAR} mission, a project led by Caltech, funded by NASA and managed by NASA/JPL, and has utilised the NuSTARDAS software package, jointly developed by the ASDC, Italy and Caltech, USA. This research has made use of observations obtained with XMM-Newton, an ESA science mission with instruments and contributions directly funded by ESA Member States and NASA. This work made use of XRT data supplied by the UK Swift Science Data Centre at the University of Leicester, UK.

\section*{Data Availability}
\label{sec:data-aval}
We used archival data of \nustar~ observatories for this work. All the models used in this work, are publicly available. Appropriate links are given in the text.



\bibliographystyle{mnras}
\bibliography{final_LLAGN} 




\appendix

\section{Spectral Variability for Non-Simultaneous Data}
\label{sec:spec-var}
We used \xmm~ observations for 12 sources in our sample. Five of these 12 \xmm~ observations were not made simultaneously with the \nustar. We checked if there is any spectral variability between the \xmm~ and \nustar~ data for non-simultaneous observations. We performed joint fitting of \xmm~ and \nustar~ data in common energy range, i.e., in $3-10$~keV range to check the variability. We used simple absorbed powerlaw model, along with a Gaussian component, for the Fe K-line. We allowed $\Gamma$ and powerlaw normalization vary freely between two data sets. We find that $\Gamma$ does not vary more than 10\% between the \xmm, and \nustar~ observations for each of the five objects, while powerlaw normalization changes. The change in the powerlaw normalization can be taken care of by using `cross normalization' factor. As $\Gamma$ did not vary more than 10\%, we used \xmm~ data with the \nustar~ data for non-simultaneous observations.

Similar to the non-simultaneous \xmm~ data, we also checked for spectral variability while adding the time-integrated BAT spectra with the \nustar~ spectra. We fitted the joint NuSTAR+BAT data in the common energy range (i.e. $15-78$ keV) with a cutoff powerlaw model. The spectral parameters ($\Gamma$ and $E_{\rm cut}$) are found to be consistent (within ~10\%) for both \nustar~ and BAT spectra in the 15--78 keV energy range.

\section{Addition of Time-Integrated BAT Spectra}
We used 105-months BAT data for the spectra analysis. The BAT spectra were added with the \nustar~ and \xmm~ or XRT data, which were obtained in short timescale ($\sim 10-100$ ks). For individual sources, there is a possibility that the spectral state of AGN could change in the BAT timescale, so it may not be appropriate to add BAT spectra with the pointed observations of \nustar~ and \xmm. However, as the current work focus on the statistical properties of LAC-AGNs, the addition of time-integrated BAT data do not change the overall findings. We checked that if spectral analysis with and without the BAT spectra and the correlation properties of various spectral parameters remains the same. The addition of the BAT data allow us to probe the spectra up to 150 keV which it improves the uncertainty of the spectral parameters.

\section{Spectral Analysis Result}
\label{sec:table-res}

\begin{table*}
\centering
\caption{Spectral Analysis Result with Model-1a and Model-1b.}
\leftline{Columns: (1) Source Name, (2) logarithm of line-of-sight of column density (\nhl), (3) photon index ($\Gamma$), (4) cut-off energy ($E_{\rm cut}$) in keV, (5) reflection fraction ($R$),}
\leftline{(6) $\chi^2$/degrees of freedom for Model-1a, (7) logarithm of line-of-sight of column density (\nhl), (8) photon index ($\Gamma$), (9) hot electron plasma temperature ($kT_{\rm e}$) }
\leftline{in keV, (10) reflection fraction ($R$), (11) $\chi^2$/degrees of freedom for Model-2.}
\leftline{Columns (2-5) represent the spectral parameters obtained from Model-1a, while columns (6-10) represent the results obtained with Model-1b.}
\begin{tabular}{ccccccccccccccc}
\hline
 & & & Model-1a & & & & & Model-1b & & \\
Object & $\log$\nhl & $\Gamma$ & $E_{\rm cut}$ & $R$ & $\chi^2$/dof &$\log$\nhl  & $\Gamma$ & $kT_{\rm e}$ &  $R$ & $\chi^2$/dof \\
       & $\log$(\pcm)& & (keV) & & & $\log$(\pcm) & & (keV) & \\
       (1) & (2) & (3) & (4) & (5) & (6) & (7) & (8) & (9) & (10) & (11) \\\hline
NGC 454E     &$23.58^{+0.06}_{-0.09} $&$ 1.80^{+0.07}_{-0.05}$&$>352               $&$ >0.91               $&231/207   &$ 23.58^{+0.06}_{-0.09} $&$1.78^{+0.05}_{-0.08}$&$>165             $&$ >0.94               $& 236/209    \\ \\
NGC 1052     &$23.16^{+0.03}_{-0.04} $&$ 1.71^{+0.04}_{-0.06}$&$254^{+152 }_{-113} $&$ >1.33               $&939/968   &$ 23.17^{+0.03}_{-0.05} $&$1.70^{+0.05}_{-0.04}$&$87 ^{+42 }_{-39} $&$ >1.44               $& 819/890    \\ \\
NGC 2110     &$22.59^{+0.02}_{-0.01} $&$ 1.68^{+0.05}_{-0.04}$&$218^{+35  }_{-47 } $&$ <0.04               $&2356/2226 &$ 22.58^{+0.02}_{-0.02} $&$1.69^{+0.05}_{-0.06}$&$73 ^{+32 }_{-15} $&$ <0.07               $& 2365/2226  \\ \\
NGC 2655     &$23.23^{+0.37}_{-0.16} $&$ 1.51^{+0.06}_{-0.08}$&$>44                $&$ >0.48               $&116/122   &$ 23.22^{+0.26}_{-0.14} $&$1.50^{+0.06}_{-0.09}$&$>23              $&$ >0.49               $& 119/122    \\ \\
NGC 3079     &$24.24^{+0.12}_{-0.07} $&$ 1.81^{+0.05}_{-0.09}$&$164^{+82  }_{-58 } $&$ >1.45               $&274/296   &$ 24.24^{+0.13}_{-0.08} $&$1.81^{+0.05}_{-0.08}$&$52 ^{+22 }_{-12} $&$ >1.44               $& 275/296    \\ \\
NGC 3147     &$22.09^{+0.06}_{-0.06} $&$ 1.90^{+0.04}_{-0.06}$&$395^{+167 }_{-44 } $&$ 0.41^{+0.34}_{-0.22}$&621/579   &$ 22.09^{+0.06}_{-0.07} $&$1.91^{+0.04}_{-0.05}$&$156^{+74 }_{-31} $&$ 0.41^{+0.35}_{-0.45}$& 619/579    \\ \\
NGC 3718     &$21.94^{+0.05}_{-0.04} $&$ 1.84^{+0.04}_{-0.06}$&$267^{+117 }_{-81 } $&$ <0.18               $&1282/1238 &$ 21.95^{+0.03}_{-0.03} $&$1.83^{+0.03}_{-0.04}$&$115^{+60 }_{-42} $&$ <0.16               $& 1285/1238  \\ \\ 
NGC 3786     &$21.83^{+0.26}_{-0.09} $&$ 1.64^{+0.05}_{-0.08}$&$186^{+102 }_{-75 } $&$ 1.04^{+0.57}_{-0.24}$&182/185   &$ 21.83^{+0.31}_{-0.13} $&$1.71^{+0.05}_{-0.06}$&$81 ^{+38 }_{-22} $&$ 1.03^{+0.56}_{-0.39}$& 181/185    \\ \\ 
NGC 3998     &$ 20.00^*              $&$ 1.86^{+0.08}_{-0.05}$&$252^{+117 }_{-82 } $&$ <0.01               $&1512/1565 &$ 20.00^{*    }_{     } $&$1.86^{+0.05}_{-0.04}$&$102^{+54 }_{-16} $&$ <0.01               $& 1519/1565  \\ \\ 
NGC 4102     &$23.85^{+0.10}_{-0.09} $&$ 1.68^{+0.05}_{-0.07}$&$201^{+72  }_{-43 } $&$ >0.44               $&401/361   &$ 23.86^{+0.09}_{-0.08} $&$1.67^{+0.06}_{-0.06}$&$72 ^{+43 }_{-2 } $&$ >0.45               $& 399/361    \\ \\ 
NGC 4258     &$23.00^{+0.10}_{-0.06} $&$ 1.77^{+0.05}_{-0.07}$&$391^{+167 }_{-121} $&$ <0.15               $&792/884   &$ 23.00^{+0.09}_{-0.06} $&$1.78^{+0.05}_{-0.06}$&$149^{+82 }_{-43} $&$ <0.19               $& 866/884    \\ \\ 
NGC 4579     &$ 20.00^*              $&$ 1.96^{+0.06}_{-0.05}$&$458^{+122 }_{-95 } $&$ <0.01               $&1647/1663 &$ 20.00^{* }_{        } $&$1.96^{+0.06}_{-0.04}$&$195^{+91 }_{-63} $&$ <0.01               $& 1649/1663  \\ \\ 
NGC 5033     &$21.67^{+0.18}_{-0.13} $&$ 1.76^{+0.05}_{-0.08}$&$201^{+168 }_{-61 } $&$ <0.18               $&1364/1379 &$ 21.71^{+0.20}_{-0.09} $&$1.76^{+0.04}_{-0.08}$&$68 ^{+18 }_{-19} $&$ <0.19               $& 1369/1369  \\ \\ 
NGC 5283     &$23.10^{+0.02}_{-0.03} $&$ 1.77^{+0.05}_{-0.08}$&$91 ^{+23  }_{-21 } $&$ 0.69^{+0.42}_{-0.12}$&428/425   &$ 23.10^{+0.02}_{-0.03} $&$1.78^{+0.05}_{-0.07}$&$45 ^{+13 }_{-0 } $&$ 0.71^{+0.45}_{-0.19}$& 428/423    \\ \\ 
NGC 5290     &$22.12^{+0.08}_{-0.09} $&$ 1.76^{+0.04}_{-0.07}$&$260^{+103 }_{-88 } $&$ 0.57^{+0.38}_{-0.43}$&515/603   &$ 22.13^{+0.14}_{-0.09} $&$1.76^{+0.05}_{-0.08}$&$104^{+61 }_{-15} $&$ 0.60^{+0.34}_{-0.38}$& 522/608    \\ \\ 
NGC 5899     &$22.94^{+0.07}_{-0.07} $&$ 1.74^{+0.06}_{-0.10}$&$115^{+43  }_{-25 } $&$ <0.52               $&326/396   &$ 22.94^{+0.06}_{-0.07} $&$1.74^{+0.05}_{-0.08}$&$45 ^{+22 }_{-12} $&$ <0.59               $& 334/396    \\ \\ 
NGC 6232     &$23.53^{+0.32}_{-0.16} $&$ 1.44^{+0.06}_{-0.10}$&$>55                $&$ >0.68               $&143/141   &$ 23.53^{+0.30}_{-0.14} $&$1.43^{+0.04}_{-0.10}$&$>21              $&$ >0.71               $& 145/141    \\ \\ 
NGC 7213     &$21.81^{+0.12}_{-0.07} $&$ 1.87^{+0.03}_{-0.05}$&$266^{+217 }_{-56 } $&$ <0.05               $&1204/1201 &$ 21.81^{+0.14}_{-0.06} $&$1.87^{+0.05}_{-0.08}$&$111^{+103}_{-32} $&$ <0.07               $& 1199/1201  \\ \\ 
NGC 7674     &$23.61^{+0.23}_{-0.11} $&$ 1.64^{+0.04}_{-0.09}$&$>47                $&$ >0.74               $&114/106   &$ 23.61^{+0.26}_{-0.11} $&$1.62^{+0.05}_{-0.09}$&$>20              $&$ >0.69               $& 118/106    \\ \\ 
Mrk 18       &$23.16^{+0.10}_{-0.10} $&$ 1.74^{+0.09}_{-0.10}$&$>277               $&$ >0.66               $&76/73     &$ 23.15^{+0.10}_{-0.09} $&$1.75^{+0.08}_{-0.04}$&$>88              $&$ >0.64               $& 76/73      \\ \\ 
Mrk 273      &$23.43^{+0.06}_{-0.03} $&$ 1.70^{+0.05}_{-0.09}$&$>238               $&$ >0.71               $&499/442   &$ 23.43^{+0.06}_{-0.04} $&$1.69^{+0.06}_{-0.09}$&$>135             $&$ >0.70               $& 502/442    \\ \\ 
ARP 102B     &$21.73^{+0.39}_{-0.14} $&$ 1.72^{+0.05}_{-0.09}$&$88 ^{+25  }_{-28 } $&$ >0.84               $&259/281   &$ 21.74^{+0.41}_{-0.16} $&$1.72^{+0.06}_{-0.09}$&$35 ^{+22 }_{-10} $&$ >0.77               $& 256/284    \\ \\ 
ESO 253--003 &$23.04^{+0.11}_{-0.14} $&$ 1.41^{+0.05}_{-0.09}$&$367^{+108 }_{-174} $&$ 0.85^{+0.35}_{-0.22}$&378/327   &$ 23.04^{+0.13}_{-0.15} $&$1.42^{+0.05}_{-0.08}$&$138^{+100}_{-41} $&$ 0.86^{+0.41}_{-0.29}$& 383/327    \\ \\ 
ESO 506--027 &$23.80^{+0.07}_{-0.04} $&$ 1.67^{+0.05}_{-0.08}$&$330^{+156 }_{-175} $&$ 0.86^{+0.41}_{-0.27}$&385/342   &$ 23.80^{+0.07}_{-0.05} $&$1.67^{+0.06}_{-0.04}$&$125^{+76 }_{-42} $&$ 0.85^{+0.45}_{-0.23}$& 379/342    \\ \\ 
HE 1136--2304&$21.08^{+0.08}_{-0.12} $&$ 1.65^{+0.05}_{-0.09}$&$231^{+75  }_{-45 } $&$ <0.15               $&2768/2559 &$ 21.08^{+0.08}_{-0.07} $&$1.65^{+0.03}_{-0.04}$&$101^{+45 }_{-21} $&$ <0.13               $& 2774/2559  \\ \\ 
IC 4518A     &$23.21^{+0.11}_{-0.06} $&$ 1.54^{+0.06}_{-0.06}$&$>41                $&$ >0.77               $&199/208   &$ 23.21^{+0.11}_{-0.07} $&$1.55^{+0.06}_{-0.07}$&$>19              $&$ >0.81               $& 198/208    \\ \\ 
IGR J11366   &$21.86^{+0.06}_{-0.07} $&$ 1.95^{+0.04}_{-0.05}$&$126^{+35  }_{-22 } $&$ 0.42^{+0.41}_{-0.34}$&495/516   &$ 21.89^{+0.05}_{-0.08} $&$1.96^{+0.08}_{-0.06}$&$59 ^{+76 }_{-15} $&$ 0.45^{+0.39}_{-0.34}$& 502/516    \\ \\ 
\hline
\end{tabular}
\label{tab:pex}
\end{table*}

\begin{table*}
\centering
\contcaption{Spectral Analysis Result with Model-1a and Model-1b.}
\leftline{Columns: (1) Source Name, (2) logarithm of line-of-sight of column density (\nhl), (3) photon index ($\Gamma$), (4) cut-off energy ($E_{\rm cut}$) in keV, (5) reflection fractio ($R$),}
\leftline{(6) $\chi^2$/degrees of freedom for Model-1a, (7) logarithm of line-of-sight of column density (\nhl), (8) photon index ($\Gamma$), (9) hot electron plasma temperature ($kT_{\rm e}$) }
\leftline{in keV, (10) reflection fraction ($R$), (11) $\chi^2$/degrees of freedom for Model-2.}
\leftline{Columns (2-5) represent the spectral parameters obtained from Model-1a, while columns (6-10) represent the results obtained with Model-1b.}
\begin{tabular}{ccccccccccccccc}
\hline
 & & & Model-1a & & & & & Model-1b & & \\
Object & $\log$\nhl & $\Gamma$ & $E_{\rm cut}$ & $R$ & $\chi^2$/dof &$\log$\nhl  & $\Gamma$ & $kT_{\rm e}$ &  $R$ & $\chi^2$/dof \\
       & $\log$(\pcm)& & (keV) & & & $\log$(\pcm) & & (keV) & \\
       (1) & (2) & (3) & (4) & (5) & (6) & (7) & (8) & (9) & (10) & (11) \\
\hline
UGC 12292    &$24.20^{+0.10}_{-0.04} $&$ 1.67^{+0.06}_{-0.09}$&$>62                $&$ >0.64                $&135/122   &$ 24.19^{+0.10}_{-0.05} $&$1.67^{+0.05}_{-0.09}$&$>40              $&$ >0.67              $& 142/122    \\ \\ 
LEDA 214543  &$22.40^{+0.12}_{-0.07} $&$ 1.72^{+0.03}_{-0.05}$&$ 98^{+45  }_{-33 } $&$ >0.77                $&465/545   &$ 22.40^{+0.14}_{-0.08} $&$1.72^{+0.03}_{-0.06}$&$39 ^{+19 }_{-11} $&$ >0.72              $& 472/545    \\ \\ 
Z367--9      &$23.26^{+0.07}_{-0.04} $&$ 1.84^{+0.05}_{-0.05}$&$170^{+108 }_{-71 } $&$ >0.64                $&234/229   &$ 23.26^{+0.08}_{-0.06} $&$1.84^{+0.05}_{-0.08}$&$52 ^{+38 }_{-21} $&$ >0.63              $& 235/229    \\ \\ 
\hline
\end{tabular}
\label{tab:pex}
\end{table*}

\begin{table*}
\centering
\caption{Spectral Analysis Result with Model-2a and Model-2b}
\leftline{Columns: (1) Source Name, (2) logarithm of line-of-sight of column density (\nhl), (3) logarithm of average column density of the obscuring materials (\nht), } 
\leftline{(4) photon index ($\Gamma$), (5) cut-off energy ($E_{\rm cut}$) in keV, (6) $\chi^2$/degrees of freedom for Model-2a, (7) logarithm of line-of-sight of column density (\nhl), (8)}
\leftline{logarithm of average column density of the obscuring materials (\nht), (9) photon index ($\Gamma$), (10) hot electron plasma temperature of the corona ($kT_{\rm e}$) in keV,}
\leftline{(11) $\chi^2$/degrees of freedom for Model-2b. }
\leftline{Columns (2-5) represent the spectral parameters obtained from Model-2a, while columns (6-10) represent the results obtained with Model-2b.}
\begin{tabular}{ccccccccccccccc}
\hline
 & & & Model-2a & & & & & Model-2b & & \\
Object & $\log$\nhl & $\log$\nht & $\Gamma$ & $E_{\rm cut}$ & $\chi^2$/dof &$\log$\nhl & $\log$\nht & $\Gamma$ & $kT_{\rm e}$ & $\chi^2$/dof \\
       & $\log$(\pcm)& log(\pcm)& & (keV) & & $\log$(\pcm) & $\log$ (\pcm) & & (keV) & \\
       (1) & (2) & (3) & (4) & (5) & (6) & (7) & (8) & (9) & (10) & (11) \\
\hline
NGC 454E        &$ 23.61^{+0.03}_{-0.03} $&$24.54^{+ 0.91}_{-1.35} $&$ 1.77^{+0.05}_{-0.05}$&$ >287               $&233/207  &$ 23.61^{+0.03}_{-0.03} $&$ 24.52^{+ 0.95}_{-1.26}$&$ 1.77^{+0.03}_{-0.01}$&$ >111            $&235/209   \\ \\   
NGC 1052        &$ 23.17^{+0.03}_{-0.03} $&$23.87^{+ 0.22}_{-0.25} $&$ 1.73^{+0.02}_{-0.03}$&$ 267^{+ 148}_{-125} $&936/968  &$ 23.17^{+0.03}_{-0.04} $&$ 23.80^{+ 0.28}_{-0.21}$&$ 1.72^{+0.02}_{-0.02}$&$  91^{+ 45}_{-33}$&814/890   \\ \\ 
NGC 2110        &$ 22.59^{+0.01}_{-0.02} $&$22.66^{+ 0.16}_{-0.31} $&$ 1.66^{+0.01}_{-0.01}$&$ 229^{+  28}_{- 28} $&2292/2226&$ 22.60^{+0.02}_{-0.01} $&$ 22.66^{+ 0.26}_{-0.52}$&$ 1.74^{+0.01}_{-0.01}$&$  89^{+ 15}_{-10}$&2320/2226 \\ \\ 
NGC 2655        &$ 23.23^{+0.39}_{-0.14} $&$23.59^{+ 0.56}_{-0.81} $&$ 1.52^{+0.05}_{-0.03}$&$  >49               $&110/122  &$ 23.23^{+0.37}_{-0.13} $&$ 23.59^{+ 0.62}_{-0.86}$&$ 1.53^{+0.04}_{-0.05}$&$  >19            $&113/122   \\ \\ 
NGC 3079        &$ 24.26^{+0.08}_{-0.06} $&$24.64^{+ 0.35}_{-0.16} $&$ 1.84^{+0.03}_{-0.04}$&$ 110^{+  52}_{- 29} $&262/296  &$ 24.27^{+0.09}_{-0.06} $&$ 24.62^{+ 0.34}_{-0.21}$&$ 1.85^{+0.06}_{-0.04}$&$  39^{+ 17}_{-10}$&268/297   \\ \\ 
NGC 3147        &$ 22.09^{+0.06}_{-0.05} $&$24.75^{+ 1.74}_{-0.88} $&$ 1.90^{+0.04}_{-0.04}$&$ 410^{+ 152}_{- 38} $&612/579  &$ 22.11^{+0.06}_{-0.05} $&$ 24.76^{+ 1.81}_{-0.92}$&$ 1.92^{+0.06}_{-0.06}$&$ 189^{+ 65}_{-29}$&610/579   \\ \\ 
NGC 3718        &$ 21.95^{+0.04}_{-0.04} $&$22.72^{+ 1.67}_{-0.72} $&$ 1.87^{+0.02}_{-0.02}$&$ 296^{+ 105}_{- 74} $&1261/1238&$ 21.96^{+0.03}_{-0.03} $&$ 22.78^{+ 1.54}_{-0.87}$&$ 1.89^{+0.01}_{-0.02}$&$ 142^{+ 65}_{-48}$&1276/1238 \\ \\ 
NGC 3786        &$ 21.84^{+0.24}_{-0.07} $&$24.71^{+ 0.68}_{-0.81} $&$ 1.73^{+0.03}_{-0.04}$&$ 192^{+ 109}_{- 82} $&164/185  &$ 21.83^{+0.20}_{-0.09} $&$ 24.72^{+ 0.74}_{-0.47}$&$ 1.73^{+0.04}_{-0.03}$&$  92^{+ 45}_{-35}$&165/185   \\ \\ 
NGC 3998        &$ 20.00^{*    }_{     } $&$22.91^{+ 0.24}_{-0.18} $&$ 1.90^{+0.04}_{-0.03}$&$ 276^{+ 143}_{- 66} $&1493/1565&$ 20.00^{*    }_{     } $&$ 22.90^{+ 0.28}_{-0.19}$&$ 1.90^{+0.02}_{-0.02}$&$ 132^{+ 92}_{-21}$&1496/1565 \\ \\ 
NGC 4102        &$ 23.85^{+0.09}_{-0.08} $&$25.51^{+25.00}_{-0.64} $&$ 1.70^{+0.03}_{-0.03}$&$ 206^{+  74}_{- 45} $&397/361  &$ 23.86^{+0.08}_{-0.08} $&$ 25.51^{+25.00}_{-0.76}$&$ 1.69^{+0.01}_{-0.03}$&$  74^{+ 53}_{- 9}$&396/261   \\ \\ 
NGC 4258        &$ 23.00^{+0.08}_{-0.06} $&$23.54^{+ 0.45}_{-0.71} $&$ 1.81^{+0.02}_{-0.02}$&$ 448^{+ 158}_{-128} $&779/884  &$ 23.00^{+0.08}_{-0.06} $&$ 23.56^{+ 0.52}_{-0.73}$&$ 1.82^{+0.04}_{-0.03}$&$ 197^{+ 66}_{-45}$&856/884   \\ \\ 
NGC 4579        &$ 20.00^{*     }_{    } $&$23.80^{+ 0.17}_{-0.12} $&$ 1.96^{+0.04}_{-0.05}$&$ 410^{+ 131}_{- 77} $&1646/1663&$ 20.00^{*    }_{     } $&$ 23.78^{+ 0.12}_{-0.15}$&$ 1.98^{+0.03}_{-0.02}$&$ 190^{+ 88}_{-58}$&1649/1663 \\ \\ 
NGC 5033        &$ 21.68^{+0.16}_{-0.10} $&$23.80^{+ 0.09}_{-0.09} $&$ 1.78^{+0.03}_{-0.03}$&$ 208^{+ 175}_{- 66} $&1363/1379&$ 21.72^{+0.13}_{-0.09} $&$ 23.80^{+ 0.10}_{-0.07}$&$ 1.82^{+0.02}_{-0.02}$&$  71^{+ 15}_{-21}$&1365/1379 \\ \\ 
NGC 5283        &$ 23.10^{+0.03}_{-0.03} $&$24.18^{+ 0.58}_{-0.34} $&$ 1.79^{+0.03}_{-0.04}$&$  82^{+  24}_{- 15} $&422/425  &$ 23.10^{+0.03}_{-0.03} $&$ 24.20^{+ 0.66}_{-0.45}$&$ 1.82^{+0.04}_{-0.04}$&$  39^{+ 14}_{- 8}$&423/423   \\ \\ 
NGC 5290        &$ 22.13^{+0.08}_{-0.08} $&$24.02^{+ 0.54}_{-0.68} $&$ 1.75^{+0.03}_{-0.04}$&$ 247^{+ 109}_{- 92} $&512/603  &$ 22.14^{+0.12}_{-0.08} $&$ 24.04^{+ 0.59}_{-0.78}$&$ 1.75^{+0.04}_{-0.03}$&$ 107^{+ 56}_{-18}$&493/608   \\ \\ 
NGC 5899        &$ 22.95^{+0.07}_{-0.07} $&$24.89^{+25.00}_{-0.49} $&$ 1.76^{+0.04}_{-0.03}$&$ 122^{+  35}_{- 22} $&306/396  &$ 22.94^{+0.07}_{-0.07} $&$ 24.91^{+25.00}_{-0.45}$&$ 1.76^{+0.03}_{-0.04}$&$  48^{+ 15}_{-10}$&308/396   \\ \\ 
NGC 6232        &$ 23.57^{+0.32}_{-0.15} $&$25.15^{+25.00}_{-0.51} $&$ 1.46^{+0.05}_{-0.04}$&$  >48               $&144/141  &$ 23.57^{+0.32}_{-0.15} $&$ 25.14^{+25.00}_{-0.59}$&$ 1.45^{+0.04}_{-0.04}$&$  >21            $&146/141   \\ \\ 
NGC 7213        &$ 21.81^{+0.14}_{-0.08} $&$23.45^{+ 0.14}_{-0.18} $&$ 1.91^{+0.06}_{-0.05}$&$ 312^{+ 262}_{- 48} $&1215/1201&$ 21.82^{+0.13}_{-0.07} $&$ 23.46^{+ 0.16}_{-0.22}$&$ 1.90^{+0.10}_{-0.10}$&$ 146^{+165}_{-16}$&1188/1201 \\ \\ 
NGC 7674        &$ 23.61^{+0.25}_{-0.10} $&$24.57^{+25.50}_{-1.02} $&$ 1.62^{+0.06}_{-0.09}$&$  >37               $&115/106  &$ 23.61^{+0.28}_{-0.11} $&$ 24.56^{+25.00}_{-0.94}$&$ 1.61^{+0.09}_{-0.08}$&$  >18            $&120/106   \\ \\ 
Mrk 18          &$ 23.16^{+0.10}_{-0.09} $&$25.01^{+25.00}_{-2.73} $&$ 1.79^{+0.04}_{-0.03}$&$ >320               $&79/73    &$ 23.15^{+0.09}_{-0.08} $&$ 25.51^{+25.00}_{-2.67}$&$ 1.78^{+0.08}_{-0.05}$&$ >88             $&81/73     \\ \\ 
Mrk 273         &$ 23.43^{+0.05}_{-0.04} $&$24.26^{+ 0.82}_{-0.46} $&$ 1.71^{+0.03}_{-0.05}$&$ >252               $&502/442  &$ 23.43^{+0.05}_{-0.04} $&$ 24.42^{+ 0.78}_{-0.52}$&$ 1.71^{+0.03}_{-0.02}$&$ >142            $&500/442   \\ \\ 
ARP 102B        &$ 21.75^{+0.47}_{-0.16} $&$24.69^{+ 0.68}_{-0.54} $&$ 1.75^{+0.04}_{-0.03}$&$  84^{+  22}_{- 14} $&251/281  &$ 21.77^{+0.59}_{-0.12} $&$ 24.65^{+ 0.76}_{-0.58}$&$ 1.76^{+0.04}_{-0.06}$&$  31^{+ 13}_{- 9}$&252/284   \\ \\ 
ESO 253--003    &$ 23.04^{+0.10}_{-0.15} $&$24.06^{+ 0.35}_{-0.25} $&$ 1.44^{+0.04}_{-0.04}$&$ 386^{+ 112}_{-205} $&381/327  &$ 23.04^{+0.10}_{-0.17} $&$ 24.08^{+ 0.29}_{-0.24}$&$ 1.45^{+0.05}_{-0.04}$&$ 148^{+106}_{-37}$&388/342   \\ \\ 
ESO 506--027    &$ 23.80^{+0.07}_{-0.04} $&$24.15^{+ 0.65}_{-0.41} $&$ 1.69^{+0.04}_{-0.02}$&$ 363^{+ 138}_{-159} $&386/342  &$ 23.80^{+0.07}_{-0.05} $&$ 24.12^{+ 0.75}_{-0.56}$&$ 1.69^{+0.04}_{-0.05}$&$ 138^{+ 78}_{-56}$&384/342   \\ \\ 
HE 1136--2304   &$ 21.08^{+0.08}_{-0.03} $&$23.15^{+ 0.15}_{-0.12} $&$ 1.67^{+0.09}_{-0.09}$&$ 256^{+  92}_{- 52} $&2754/2559&$ 21.11^{+0.07}_{-0.06} $&$ 23.20^{+ 0.21}_{-0.17}$&$ 1.67^{+0.05}_{-0.03}$&$ 116^{+ 44}_{-38}$&2752/2559 \\ \\ 
IC 4518A        &$ 23.21^{+0.10}_{-0.07} $&$24.76^{+ 0.53}_{-0.30} $&$ 1.55^{+0.04}_{-0.09}$&$  >35               $&198/208  &$ 23.20^{+0.12}_{-0.07} $&$ 24.73^{+ 0.65}_{-0.45}$&$ 1.56^{+0.04}_{-0.05}$&$  >15            $&195/208   \\ \\ 
IGR J11366      &$ 21.88^{+0.07}_{-0.04} $&$24.30^{+ 0.32}_{-0.25} $&$ 1.94^{+0.06}_{-0.06}$&$ 109^{+  32}_{- 19} $&487/516  &$ 21.89^{+0.06}_{-0.04} $&$ 24.27^{+ 0.36}_{-0.24}$&$ 1.96^{+0.06}_{-0.06}$&$  58^{+ 80}_{-13}$&495/516   \\ \\ 
\hline
\end{tabular}
\leftline{$^*$ fixed during analysis.}
\label{tab:borus}
\end{table*}

\begin{table*}
\centering
\contcaption{Spectral Analysis Result with Model-2a \& Model-2b.}
\leftline{Columns: (1) Source Name, (2) logarithm of line-of-sight of column density (\nhl), (3) logarithm of average column density of the obscuring materials (\nht), } 
\leftline{(4) photon index ($\Gamma$), (5) cut-off energy ($E_{\rm cut}$) in keV, (6) $\chi^2$/degrees of freedom for Model-2a, (7) logarithm of line-of-sight of column density (\nhl), (8)}
\leftline{logarithm of average column density of the obscuring materials (\nht), (9) photon index ($\Gamma$), (10) hot electron plasma temperature of the corona ($kT_{\rm e}$) in keV,}
\leftline{(11) $\chi^2$/degrees of freedom for Model-2b. }
\leftline{Columns (2-5) represent the spectral parameters obtained from Model-2a, while columns (6-10) represent the results obtained with Model-2b.}
\begin{tabular}{ccccccccccccccc}
\hline
 & & & Model-2a & & & & & Model-2b & & \\
Object & $\log$\nhl & $\log$\nht & $\Gamma$ & $E_{\rm cut}$ & $\chi^2$/dof &$\log$\nhl & $\log$\nht & $\Gamma$ & $kT_{\rm e}$ & $\chi^2$/dof \\
       & $\log$(\pcm)& log(\pcm)& & (keV) & & $\log$(\pcm) & $\log$ (\pcm) & & (keV) & \\
       (1) & (2) & (3) & (4) & (5) & (6) & (7) & (8) & (9) & (10) & (11) \\
\hline
UGC 12282       &$ 24.20^{+0.08}_{-0.05} $&$24.82^{+ 0.45}_{-0.59} $&$ 1.69^{+0.04}_{-0.06}$&$ >57                $&137/122  &$ 24.21^{+0.10}_{-0.06} $&$ 24.83^{+ 0.75}_{-0.85}$&$ 1.68^{+0.03}_{-0.05}$&$  >39            $&144/122   \\ \\
LEDA214543      &$ 22.39^{+0.10}_{-0.07} $&$24.56^{+ 0.45}_{-0.36} $&$ 1.71^{+0.02}_{-0.04}$&$ 106^{+  52}_{- 29} $&452/545  &$ 22.39^{+0.10}_{-0.07} $&$ 24.56^{+ 0.48}_{-0.35}$&$ 1.71^{+0.05}_{-0.05}$&$  36^{+ 21}_{-10}$&467/545   \\ \\
Z367--9         &$ 23.27^{+0.07}_{-0.04} $&$24.10^{+ 0.72}_{-0.46} $&$ 1.86^{+0.03}_{-0.02}$&$ 175^{+ 114}_{- 79} $&232/229  &$ 23.27^{+0.08}_{-0.05} $&$ 24.05^{+ 0.68}_{-0.52}$&$ 1.88^{+0.04}_{-0.05}$&$  56^{+ 41}_{-25}$&233/229   \\ \\
\hline
\end{tabular}
\leftline{$^*$ fixed during analysis.}
\label{tab:borus}
\end{table*}

\begin{table*}
\centering
\caption{Important Parameters}
\leftline{Columns: (1) Source Name, (2) logarithm of the bolometric luminosity, (3) logarithm of the Eddington ratio, (4) dimensionless temperature,}
\leftline{(5) optical depth of the hot electron plasma, (6) compactness parameter, (7) number of scattering of the seed photons in the Compton cloud,}
\leftline{(8) logarithm of the jet luminosity.} 
\begin{tabular}{ccccccccccccc}
\hline
Object & $\log~L_{\rm bol}$ & $\log \lambda_{\rm Edd}$ &  $\theta$ & $\tau_{\rm e}$ & $l$ & $N_{\rm S}$ & $\log~L_{\rm jet}$ \\
  & $\log$~(\eps)& &  & & &  & $\log$~(\eps)\\
    (1) & (2) & (3) & (4) & (5) & (6) & (7)  \\
\hline
NGC 454E        &$ 43.05\pm0.02 $&$-3.584 \pm0.472 $&$ >0.23               $&$ <1.33                $&$ 0.60^{+1.17}_{-0.40}   $&$  >2          $&$ 44.23\pm 0.03$\\ \\
NGC 1052        &$ 42.63\pm0.01 $&$-4.444 \pm0.304 $&$ 0.18^{+0.09}_{-0.06}$&$ 1.42^{+0.41}_{-0.61} $&$ 0.08^{+0.08}_{-0.04}   $&$  6^{+2}_{- 2}$&$ 44.78\pm 0.08$\\ \\
NGC 2110        &$ 43.37\pm0.02 $&$-4.124 \pm0.121 $&$ 0.17^{+0.03}_{-0.02}$&$ 1.41^{+0.17}_{-0.15} $&$ 0.17^{+0.05}_{-0.04}   $&$  6^{+1}_{- 1}$&$ 44.66\pm 0.04$\\ \\
NGC 2655        &$ 41.97\pm0.01 $&$-3.844 \pm0.208 $&$ >0.03               $&$ <7.72                $&$ 0.33^{+0.20}_{-0.13}   $&$ >23          $&$ 43.78\pm 0.03$\\ \\
NGC 3079        &$ 42.54\pm0.02 $&$-3.844 \pm0.321 $&$ 0.08^{+0.03}_{-0.02}$&$ 2.28^{+0.57}_{-0.63} $&$ 0.33^{+0.36}_{-0.17}   $&$  9^{+2}_{- 2}$&$ 44.37\pm 0.04$\\ \\
NGC 3147        &$ 42.68\pm0.02 $&$-4.224 \pm0.122 $&$ 0.37^{+0.13}_{-0.06}$&$ 0.62^{+0.16}_{-0.15} $&$ 0.14^{+0.04}_{-0.03}   $&$  2^{+1}_{- 1}$&$ 44.25\pm 0.03$\\ \\
NGC 3718        &$ 41.52\pm0.02 $&$-6.124 \pm0.122 $&$ 0.28^{+0.13}_{-0.09}$&$ 0.82^{+0.23}_{-0.35} $&$ 0.002^{+0.001}_{-0.001}$&$  3^{+3}_{- 1}$&$ 42.40\pm 0.09$\\ \\
NGC 3786        &$ 42.18\pm0.02 $&$-3.464 \pm0.118 $&$ 0.18^{+0.09}_{-0.07}$&$ 1.39^{+0.42}_{-0.66} $&$ 0.79^{+0.25}_{-0.19}   $&$  6^{+3}_{- 2}$&$ 43.55\pm 0.05$\\ \\
NGC 3998        &$ 42.30\pm0.02 $&$-5.744 \pm0.241 $&$ 0.26^{+0.14}_{-0.04}$&$ 0.86^{+0.28}_{-0.15} $&$ 0.004^{+0.003}_{-0.002}$&$  3^{+1}_{- 2}$&$ 43.43\pm 0.05$\\ \\
NGC 4102        &$ 42.42\pm0.01 $&$-4.444 \pm0.344 $&$ 0.14^{+0.09}_{-0.02}$&$ 1.72^{+0.55}_{-0.24} $&$ 0.08^{+0.10}_{-0.04}   $&$  7^{+1}_{- 2}$&$ 43.67\pm 0.02$\\ \\
NGC 4258        &$ 41.39\pm0.03 $&$-4.294 \pm0.382 $&$ 0.39^{+0.13}_{-0.09}$&$ 0.68^{+0.17}_{-0.19} $&$ 0.12^{+0.16}_{-0.07}   $&$  2^{+1}_{- 1}$&$ 43.59\pm 0.03$\\ \\
NGC 4579        &$ 42.65\pm0.02 $&$-3.564 \pm0.121 $&$ 0.37^{+0.17}_{-0.11}$&$ 0.58^{+0.18}_{-0.23} $&$ 0.63^{+0.20}_{-0.15}   $&$  2^{+1}_{- 1}$&$ 43.75\pm 0.04$\\ \\
NGC 5033        &$ 42.15\pm0.02 $&$-3.824 \pm0.373 $&$ 0.14^{+0.03}_{-0.04}$&$ 1.52^{+0.24}_{-0.49} $&$ 0.35^{+0.46}_{-0.20}   $&$  6^{+2}_{- 1}$&$ 42.39\pm 0.05$\\ \\
NGC 5283        &$ 43.01\pm0.01 $&$-3.974 \pm0.311 $&$ 0.08^{+0.03}_{-0.02}$&$ 2.35^{+0.54}_{-0.51} $&$ 0.25^{+0.25}_{-0.13}   $&$  9^{+2}_{- 2}$&$ 43.48\pm 0.05$\\ \\
NGC 5290        &$ 42.83\pm0.02 $&$-3.044 \pm0.334 $&$ 0.21^{+0.11}_{-0.04}$&$ 1.21^{+0.39}_{-0.24} $&$ 2.09^{+2.37}_{-1.11}   $&$  5^{+1}_{- 2}$&$ 43.74\pm 0.04$\\ \\
NGC 5899        &$ 43.16\pm0.02 $&$-3.614 \pm0.334 $&$ 0.09^{+0.03}_{-0.02}$&$ 2.16^{+0.44}_{-0.50} $&$ 0.56^{+0.64}_{-0.30}   $&$  9^{+2}_{- 2}$&$ 43.78\pm 0.04$\\ \\
NGC 6232        &$ 42.48\pm0.01 $&$-3.064 \pm0.527 $&$ >0.04               $&$ <5.98                $&$ 1.99^{+4.76}_{-1.40}   $&$ >23          $&$ 44.33\pm 0.03$\\ \\
NGC 7213        &$ 43.06\pm0.01 $&$-3.044 \pm0.111 $&$ 0.29^{+0.24}_{-0.03}$&$ 0.79^{+0.36}_{-0.15} $&$ 2.09^{+0.60}_{-0.47}   $&$  3^{+1}_{- 2}$&$ 44.27\pm 0.06$\\ \\
NGC 7674        &$ 43.15\pm0.04 $&$-4.144 \pm0.141 $&$ >0.04               $&$ <6.02                $&$ 0.17^{+0.06}_{-0.05}   $&$ >20          $&$ 45.75\pm 0.03$\\ \\
Mrk 18          &$ 42.82\pm0.02 $&$-3.144 \pm0.318 $&$ >0.17               $&$ <1.98                $&$ 1.66^{+1.80}_{-0.86}   $&$ >2           $&$ 43.94\pm 0.04$\\ \\
\hline
\end{tabular}
\label{tab:corona}
\end{table*}

\begin{table*}
\centering
\contcaption{Important Parameters}
\begin{tabular}{ccccccccccccc}
\hline
Object & $\log~L_{\rm bol}$ & $\log \lambda_{\rm Edd}$ &  $\theta$ & $\tau_{\rm e}$ & $l$ & $N_{\rm S}$ & $\log~L_{\rm jet}$ \\
  & $\log$~(\eps)& &  & & &  & $\log$~(\eps)\\
    (1) & (2) & (3) & (4) & (5) & (6) & (7)  \\
\hline
Mrk 273         &$ 43.68\pm0.02 $&$-3.454 \pm0.062 $&$ >0.28               $&$ <1.11                $&$ 0.81^{+0.12}_{-0.11} $&$  >2          $&$ 45.69\pm 0.03$\\ \\
ARP 102B        &$ 43.58\pm0.02 $&$-3.454 \pm0.363 $&$ 0.06^{+0.03}_{-0.02}$&$ 2.92^{+0.72}_{-0.98} $&$ 0.81^{+1.05}_{-0.45} $&$ 12^{+4}_{- 3}$&$ 45.28\pm 0.03$\\ \\
ESO 253--003    &$ 43.83\pm0.02 $&$-4.124 \pm0.124 $&$ 0.27^{+21}_{-0.05}  $&$ 1.57^{+1.10}_{-0.42} $&$ 0.17^{+0.06}_{-0.04} $&$  8^{+2}_{- 4}$&$ 46.44\pm 0.14$\\ \\
ESO 506--027    &$ 44.07\pm0.02 $&$-3.034 \pm0.314 $&$ 0.27^{+0.15}_{-0.11}$&$ 1.07^{+0.37}_{-0.63} $&$ 2.13^{+2.22}_{-1.09} $&$  4^{+3}_{- 2}$&$ 45.02\pm 0.04$\\ \\
HE 1136--2304   &$ 44.25\pm0.02 $&$-3.254 \pm0.121 $&$ 0.23^{+0.09}_{-0.05}$&$ 1.26^{+0.33}_{-0.32} $&$ 1.29^{+0.41}_{-0.31} $&$  5^{+1}_{- 1}$&$ 44.28\pm 0.04$\\ \\
IC 4518A        &$ 43.37\pm0.02 $&$-3.304 \pm0.125 $&$ >0.03               $&$ <6.67                $&$ 1.15^{+0.36}_{-0.28} $&$ >26          $&$ 45.26\pm 0.10$\\ \\
IGR J11366--3602&$ 43.42\pm0.02 $&$-3.484 \pm0.121 $&$ 0.11^{+0.16}_{-0.03}$&$ 1.53^{+0.81}_{-0.42} $&$ 0.76^{+0.24}_{-0.18} $&$  6^{+2}_{- 3}$&$  -           $\\ \\
UGC 12282       &$ 42.65\pm0.02 $&$-5.264 \pm0.371 $&$ >0.08               $&$ <2.91                $&$ 0.02^{+0.01}_{-0.01} $&$  >12         $&$ 43.89\pm 0.02$\\ \\
LEDA214543      &$ 44.31\pm0.02 $&$-3.634 \pm0.338 $&$ 0.07^{+0.04}_{-0.02}$&$ 2.79^{+0.88}_{-0.87} $&$ 0.54^{+0.64}_{-0.29} $&$ 11^{+3}_{- 4}$&$ 43.72\pm 0.05$\\ \\
Z367--9         &$ 43.56\pm0.01 $&$-4.374 \pm0.332 $&$ 0.11^{+0.08}_{-0.05}$&$ 1.70^{+0.63}_{-1.02} $&$ 0.10^{+0.11}_{-0.05} $&$  7^{+4}_{- 2}$&$ 43.80\pm 0.10$\\ \\
\hline
\end{tabular}
\label{tab:corona}
\end{table*}

\section{Corner Plot}
\label{sec:mcmc}

\begin{figure*}
\centering
\includegraphics[width=8.5cm]{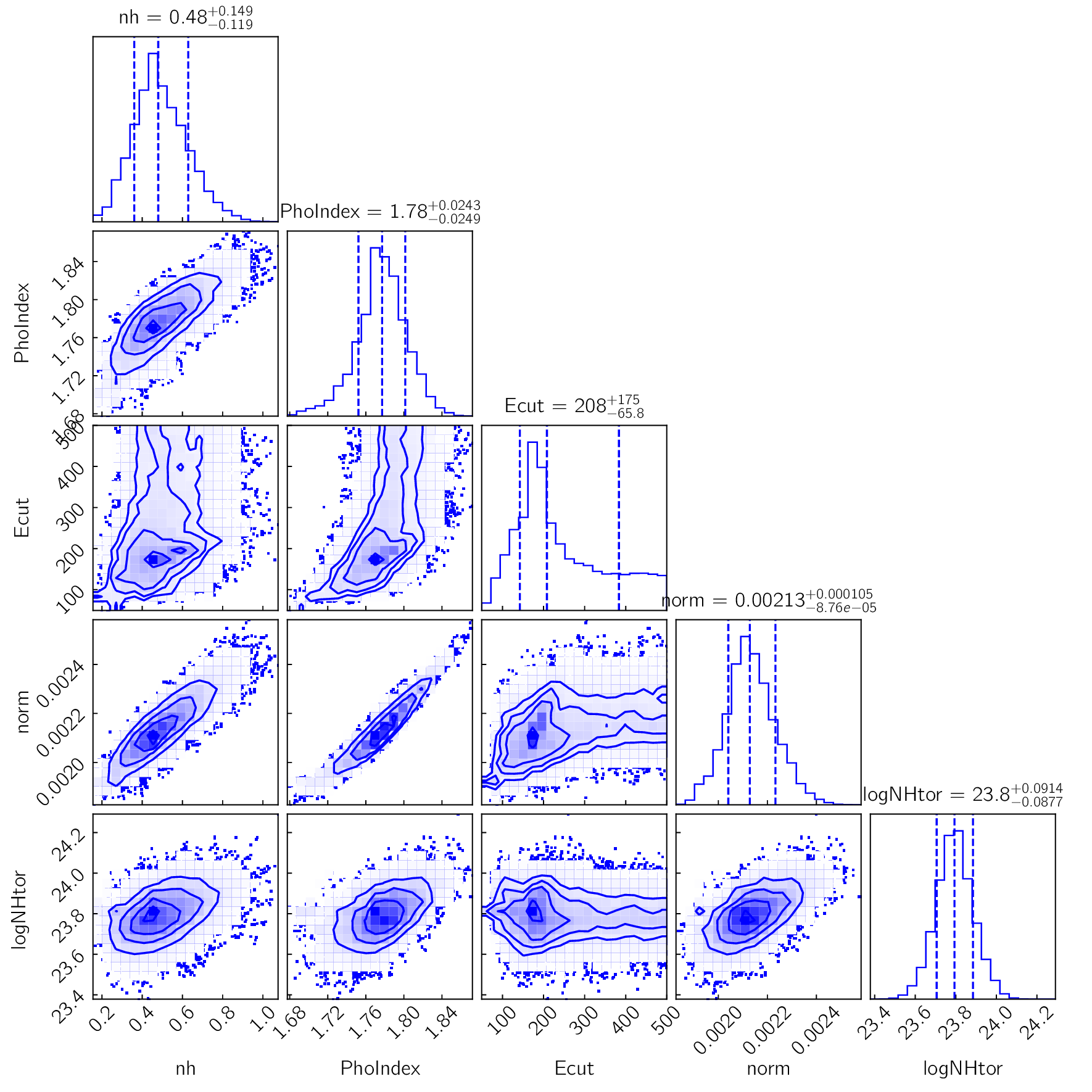}
\includegraphics[width=8.5cm]{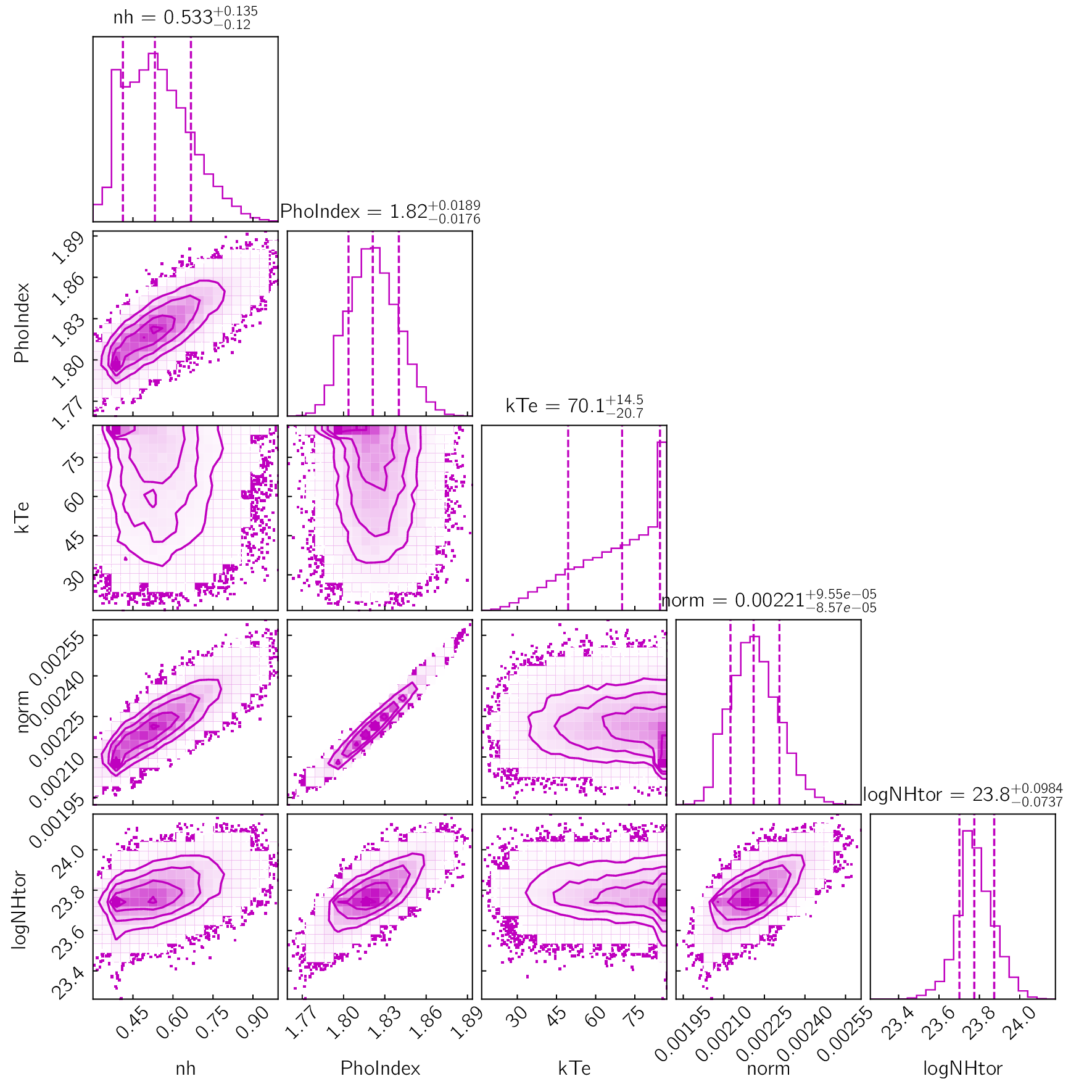}
\caption{Posterior distribution of the spectral parameters obtained from the MCMC analysis with the Model-1 and Model-2, in the left and right panel, respectively. Plotting was performed using corner plot \citep{corner}. Central dashed lines correspond to the peak values whereas $1 \sigma$ confidence levels are represented by dashed lines on either sides.}
\label{fig:mcmc}
\end{figure*}

\bsp	
\label{lastpage}
\end{document}